\documentclass[sn-mathphys-num]{sn-jnl}
\usepackage{graphicx}%
\usepackage{multirow}%
\usepackage{amsmath,amssymb,amsfonts,bbm,bm}%
\usepackage{amsthm}%
\usepackage{mathrsfs}%
\usepackage[title]{appendix}%
\usepackage{xcolor}%
\usepackage{textcomp}%
\usepackage{manyfoot}%
\usepackage{booktabs}%
\usepackage{algorithm}%
\usepackage{algorithmicx}%
\usepackage{algpseudocode}%
\usepackage{listings}%
\usepackage{array}
\usepackage{textcomp}
\usepackage{stfloats}
\usepackage{url}
\usepackage{verbatim}
\usepackage{float}
\usepackage{hyperref}
\usepackage{subcaption}

\usepackage{multirow}
\usepackage{makecell}
\theoremstyle{thmstyleone}%
\theoremstyle{thmstyletwo}%
\theoremstyle{thmstylethree}%
\makeatletter
\newcommand{\tmb}[3]{%
  \mathrel{%
    \vbox{\offinterlineskip\m@th
      \ialign{%
        \hfil##\hfil\cr
        $\scriptscriptstyle#1\mathstrut$\cr
        \noalign{\vspace{0.3ex}}
        \vtop{%
          \ialign{%
            \hfil##\hfil\cr
            $#2$\cr
            $\scriptscriptstyle#3\mathstrut$\cr
          }%
        }\cr
      }%
    }%
  }%
}
\makeatother
\def\beq{\begin{equation}}
\def\eeq{\end{equation}}
\def\bea{\begin{eqnarray}}
\def\eea{\end{eqnarray}}
\def\ba{\begin{array}}
\def\ea{\end{array}}

\def\bitem{\begin{itemize}}
\def\eitem{\end{itemize}}
\def\ben{\begin{enumerate}}
\def\een{\end{enumerate}}
\def\bss{\begin{singlespace}}
\def\ess{\end{singlespace}}

\def\eg{{\it e.g., \/}}

\def\ie{{\it i.e.,\ \/}}

\definecolor{bgrd}{rgb}{1,1,1}
\definecolor{gray}{rgb}{0.5,0.5,0.5}
\definecolor{dkr}{rgb}{0.7,0.1,0.2}
\definecolor{dkb}{rgb}{0.1,0.1,0.8}

\newcommand{\mbbE}{\mathbb{E}}

\newcommand{\Pmsc}{\mathscr{P}}
\newcommand{\Qmsc}{\mathscr{Q}}

\def\nubf{\hbox{\boldmath$\nu$\unboldmath}}

\def\vbf{{\bm v}}

\def\xbf{{\bm x}}

\def\xbf{{\bm x}}

\def\Gc{{\cal G}}
\def\Hc{{\cal H}}

\def\Kc{{\cal K}}

\def\Uc{{\cal U}}

\begin{document}
\title[Article Title]{Grid Monitoring with Synchro-Waveform and AI Foundation Model Technologies}
\author*[1]{\fnm{Lang} \sur{Tong}}\email{LT35@cornell.edu}
\author[1]{\fnm{Xinyi} \sur{Wang}}\email{xw555@cornell.edu}
\author[1]{\fnm{Qing} \sur{Zhao}}\email{qz16@cornell.edu}
\affil*[1]{\orgdiv{School of Electrical and Computer Engineering},  \orgname{Cornell University} \\[0.3em] \orgaddress{ \city{Ithaca}, \postcode{14850}, \state{NY}, \country{USA}}}
\abstract{\textbf{Purpose:}
This article advocates for the development of a next-generation grid monitoring and control system designed for future grids dominated by inverter-based resources. Leveraging recent progress in generative artificial intelligence (AI), machine learning, and networking technology, we develop a physics-based AI foundation model with high-resolution synchro-waveform measurement technology to enhance grid resilience and reduce economic losses from outages.

\textbf{Methods and Results:}
The proposed framework adopts the AI Foundation Model paradigm, where a generative and pre-trained (GPT) foundation model extracts physical features from power system measurements, enabling adaptation to a wide range of grid operation tasks. Replacing the large language models used in popular AI foundation models, this approach is based on the Wiener-Kallianpur-Rosenblatt innovation model for power system time series, trained to capture the physical laws of power flows and sinusoidal characteristics of grid measurements. The pre-trained foundation model causally extracts sufficient statistics from grid measurement time series for various downstream applications, including anomaly detection, over-current protection, probabilistic forecasting, and data compression for streaming synchro-waveform data. Numerical simulations using field-collected data demonstrate significantly improved fault detection accuracy and detection speed.

\textbf{Conclusion:}
The future grid will be rich in inverter-based resources, making it highly dynamic, stochastic, and low inertia. This work underscores the limitations of existing Supervisory-Control-and-Data-Acquisition and Phasor-Measurement-Unit monitoring systems and advocates for AI-enabled monitoring and control with high-resolution synchro-waveform technology to provide accurate situational awareness, rapid response to faults, and robust network protection.
}

\keywords{Generative AI, AI Foundation Model, Continuous-point-on-wave measurements, Synchro-waveform measurements, Power system monitoring and control, Power signal compression, Fault detection, and Over-current Protection.}

\maketitle

\section{Introduction}
Over the past decades, the power grid has undergone unprecedented changes, evolving from a system dominated by high-inertia synchronous generators dispatched to meet predictable demands to one with substantial non-dispatchable, stochastic, and inverter-based renewable resources (IBRs) such as wind, solar, and battery storage devices. Operating at microsecond timescales, these resources reduce system inertia, making the grid more stochastic and dynamic, frequently operating near its limits.

Externally, the accelerated climate-related events place enormous stress on the aging grid. Outage events between 2011 and 2021 increased by 78\% compared to the previous decade \cite{ClimatMattes:2022}, resulting in an average annual economic loss exceeding 150 billion dollars \cite{Hussain:19}. While not all outages can be prevented, this raises a critical question: how many outages can be mitigated, load sheds restored quickly, and economic impacts minimized by advanced computing, networking, and artificial intelligence (AI)?

This paper advocates for a next-generation AI-powered grid monitoring system built on synchro-waveform measurement technology, enabling high-resolution and high-fidelity monitoring beyond the capabilities of state-of-the-art SCADA\footnote{Supervisory Control and Data Acquisition} and Phasor Measurement Unit (PMU)  technologies. To achieve this, we introduce a physics-based {\em AI Foundation Model} framework, augmenting the popular Large Language Model (LLM) used in most AI foundation models for natural language processing with the classic Wiener-Kallianpur-Rosenblatt {\em Innovation Representation Model} (IRM) \cite{Wiener:58Book,Masani:66BAMS} for power system signal processing for monitoring, control, and real-time decision-making.

\subsection{From Synchro-phasor to Synchro-waveform Technology}
The synchro-waveform technology generates direct and synchronously sampled current and voltage waveform measurements, and a device that produces such direct measurements is called Waveform Measurement Units (WMU) \cite{Mohsenian-Rad&Xu:23}.   WMUs are ubiquitous in modern grids. Digital sensors, intelligent electronic devices (IEDs), PMUs, digital relays, and digital fault recorders all inherently generate waveform-sampled data with sufficiently fast sampling rates to capture the highly dynamic waveforms of grid events before converting them to statistics (such as phasors) for monitoring tasks. If the next-generation monitoring system is based on high-resolution synchro-waveform measurements, compelling arguments are needed for adopting new WMU in favor of the mature PMU and  SCADA technologies. What can WMU measurements reveal that SCADA and PMU measurements cannot? Do the benefits of WMU data justify the costs of new devices and infrastructure investments?

The PMU technology was introduced over three decades ago as a transformative successor to the SCADA technology for wide-area situational awareness in power systems. Despite its potential, PMU deployment in U.S. utilities remains limited, with only a few thousand units installed to date---insufficient for high-resolution monitoring across the operational footprint of even small to medium-sized system operators. After decades of substantial global investment, primarily driven by governmental initiatives and extensive academic and industrial research, the technology has yet to achieve its anticipated impact. In light of this stagnation, it is difficult to envision PMU technology making a transformative impact in addressing emerging challenges in an IBR-dominated power grid.

\begin{figure}[h]
\centerline{\includegraphics[scale=0.6]{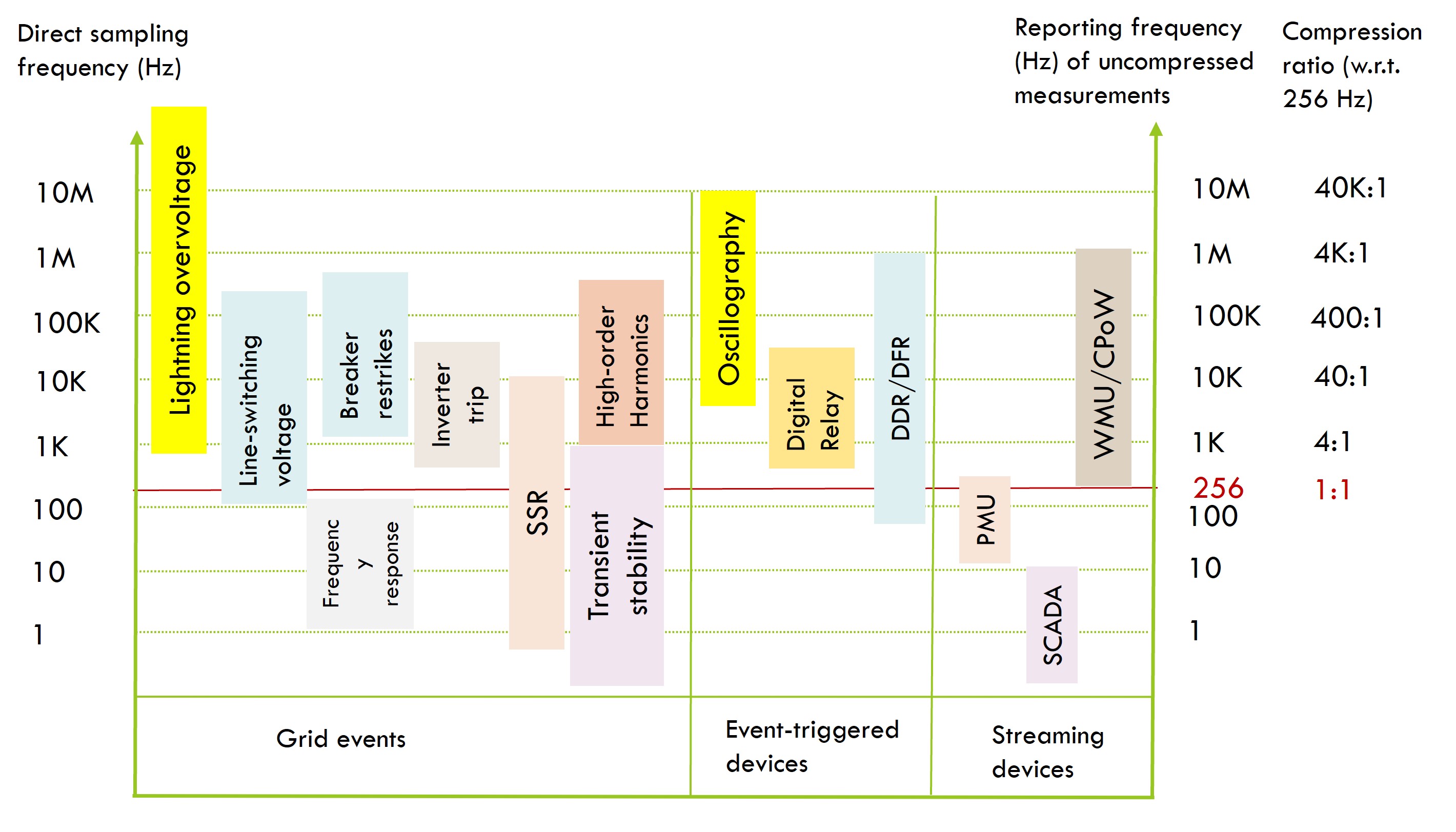}}
\caption{\small Frequency contents of various grid events, each represented by a bar covering the range of the necessary Nyquist sampling rates on the left axis.  The right axis shows the necessary reporting frequency of uncompressed measurement and the required compression ratio (relative to the 256 Hz PMU reporting rate.)  This Figure is adapted from \cite{Wang&Liu&Tong:21TPS,Silverstein&Follum:20,Anderson&Agrawal&Ness:99,Perez:10}.}
\label{fig:events}
\end{figure}

The barrier to entry for the PMU technology is twofold. As a technology, PMU has a limited reporting rate, capturing only a small fraction of significant grid events, as illustrated in Fig.\ref{fig:events}. A PMU extracts current and voltage phasors associated with the fundamental frequency of 50 or 60 Hz, filtering out frequency components beyond a narrow band around the fundamental frequency. Consequently, PMU measurements are insensitive to high-order and intra-harmonics of many critical grid events shown in Fig.\ref{fig:events}, inadequate to monitor critical dynamics during network stress while generating excessive data under normal operating conditions. The so-called chicken-and-egg economic barrier also exists: insufficient monitoring resolution and limited network coverage make PMUs ineffective for practical wide-area monitoring of impactful events, which limits the development of compelling use cases. The lack of convincing applications limits investments and deployments necessary to unlock the technology’s full potential.

Two key arguments are often raised against adopting the higher-resolution synchro-waveform technology. The first is the cost of infrastructure investments and bandwidth requirements to stream these data. However, this argument is increasingly unconvincing in light of advancements in compression, communications, and networking technologies over the last two decades. Since the invention of PMU, communication rates have increased by nearly six orders of magnitude. Today’s communication networks can seamlessly handle high-fidelity entertainment, real-time streaming of live events, and control of autonomous systems such as drones and robots. The notion that streaming synchro-waveform data is infeasible is no longer convincing.

The second argument pertains to the perceived benefits of using synchro-waveform data. The PMU technology has been criticized for contributing to a ‘data deluge’ during normal grid operation, and the WMU technology would generate 10 to 100 times more data, which appears to be a considerable waste of resources devoted to relatively rare occasions. A counterargument is that the grid is not always in a normal or failure state; often, the transition from normal to failure involves sequences of incipient faults that are sporadic and sometimes self-clearing. Detecting such faults requires high-resolution measurements. Discounting the potential of high-resolution measurement technology stifles progress toward breakthrough technologies critical for grid reliability and resiliency.

More convincingly, the challenge of managing data deluge lies in the technology that effectively sifts through large datasets to identify early signs of instability and actionable information. A practical solution to balance the need for high-resolution data and avoiding data overload is the hierarchical push-pull networking: push data from WMUs to the control center when local events are detected, pull data from WMUs when global assessment is necessary. The mathematical foundations and practical implementations of such protocols for content delivery networking have long existed in the literature. See, \eg \cite{Cybenko:99Bkchap,Pathan&etal:08Bkchap}.

While the WMU technology has not been deployed and effective data analytic techniques missing, the benefits of using synchro-waveform have been well articulated for almost a decade, see \eg \cite{Wischkaemper:15TSG,Bastos&etal:19PESGM,Carroll19NAPSI,Rahmatian:19,Izadi&Mohsenian-Rad:20IGST,Xu&etal:22TPD}. Recent reviews of synchro-waveform measurement technology  \cite{Xu&etal:22TPD,Mohsenian-Rad&Xu:23} have articulated broad visions for its use in monitoring, protection, and control. This article aligns with these visions, offering a possible architectural design within the modern AI foundation model paradigm and addressing the data analytic challenges posed by highly stochastic measurements.

\subsection{Scope, Contributions, and Organization}
This paper develops key components of an AI-enhanced WMU-based grid monitoring system, including (i) a novel foundation model architecture using the {\em innovation representation model} (IRM) for synchro-waveform time series, (ii) the pre-training of the IRM-based foundation model, and (iii) foundation model adaptations for anomaly detection, over-current protection, and synchro-waveform data compression.

The main contributions of this work are as follows.
\ben
\item {\bf An IRM-based foundation model architecture:} A fundamental challenge in compressing and performing statistical inference on synchro-waveform streaming is the unknown probability models of WMU data. To address this, we propose the {\em innovation representation model} (IRM) based on the classic work of representation of random processes by Norbert Wiener and Gopinath Kallianpur \cite{Wiener:58Book,Masani:66BAMS}  and Murray Rosenblatt \cite{Rosenblatt:59}. IRM provides a unified data processing foundation for all statistical decision tasks, including the monitoring, protection, probabilistic forecasting, and compression applications considered here.

\item {\bf Rapid detection of faults and novel trends:}
 We introduce a sequential detection method that identifies anomalies and novel trends in WMU data streams. Unlike conventional methods that rely on direct comparisons of current and voltage phasors to nominal values, our approach operates in the innovation feature space. A novel kernel technique based on a sequential adaptation of the classic Neyman Smooth Test (NST) ensures rapid detection with high confidence.
 \item {\bf Application to over-current protection:}
 The proposed anomaly detection method is applied to over-current protection in distribution systems, addressing critical challenges such as protection blinding and sympathetic tripping caused by stochastic distributed generation. We demonstrate significant improvements in both detection accuracy and speed.

 \item {\bf Data compression for synchro-waveform streaming:} We develop a novel adaptive subband compression architecture that integrates innovation representation for compressing WMU data. By dividing WMU measurements into harmonic and inter-harmonic subbands, the proposed compression approach compresses and communicates selected active subbands based on their levels of innovation features. This method is crucial for enabling bandwidth-efficient, dynamic, and interactive operations among intelligent electronic devices (IEDs), substations, and control centers, especially during severe grid events.
 \een

This paper is organized as follows. Sec.~\ref{sec:FM} introduces an AI foundation model approach to power system monitoring and control based on the innovation representation of power system measurements. We discuss the architecture of the proposed foundation model and foundation model pre-training using WMU data for downstream foundation model adaptations. Three foundation model adaptations are then discussed in subsequent sections.
Sec.~\ref{sec:anomaly} presents the adaptation of the proposed foundation model for anomaly detection. A novel sequential detection scheme based on Neyman's smooth test is derived to achieve fast detection speed and high accuracy. Sec.~\ref{sec:overcurrent} presents the foundation model adaptation for over current protection. Sec.~\ref{sec:compression} presents the foundation model adaptation for the event-driven compression of WMU measurements.

Notations used in this work are standard in statistical learning and power system literatures. Vectors, matrices, and sequences are usually in boldface.
When considering a segment of a sequence $\xbf$, $\xbf_{t_1:t_2}$ denotes the segment from $t_1$ to $t_2$, \ie $\xbf_{t_1:t_2}=(x_{t_1}, \cdots, x_{t_2})$. When necessary to distinguish a random variable and its realization, a random variable is capitalized and its realization in small letter.   For two random variables $X$ and $Y$, $X\stackrel{\mbox{\sf\small a.s.}}{=}Y$ means the two equal almost surely and $X\stackrel{\mbox{\sf\small d}}{=}Y$ equal in distribution.  Table~\ref{tab:notation} gives a list of designated symbols and key abbreviations.

\begin{table}[htbp]
    \centering
    \caption{Major Nomenclatures and Mathematical Symbols.}
    \label{tab:notation}
        \setlength\tabcolsep{1pt}

    \begin{tabular}{l l}
    \hline
    DG/SDG & Distributed generation/Stochastic distributed generation.\\
    IBR & Inverter-based resources.\\
    IID: & Independent and Identically Distributed\\
IID-Uniform & IID random variables with uniformly distributed marginals\\
    LLM/IRM: & Large Language Model/Innovation Representation Model.\\
    NST/SNST: & Neyman's Smooth Test/Sequential Neyman's Smooth Test.\\
    PMU/WMU: & Phasor Measurement Unit/Waveform Measurement Unit.\\\hline
        $(x_t)$: & a sequence of WMU measurements.\\
        $(\nu_t)$: & an innovation sequence. \\
        $(u_t)$: & the IID sequence with the uniformly distributed marginal.\\
        $(\hat{x}_t)$: & the reconstructed $(x_t)$ at the output by the foundation model autoencoder.\\
        $G$: & the innovation encoder function.\\
        $H$: & the innovation decoder function.\\
        $G_\theta$: & the neural network representation of $G$ parameterized by $\theta$.\\
        $H_\eta$: & the neural network representation of $H$ parameterized by $\eta$.\\
        $D_\gamma$: & the neural network representation of innovation discriminator between $(\nu_t)$ and $(u_t)$.\\
        $\Uc[0,1]$: & the continuous univariate uniform distribution on $[0,1]$.\\\hline
    \end{tabular}
\end{table}

\section{A Foundation Model for Synchro-waveform Data}
\label{sec:FM}
AI Foundation Model, broadly defined, is a learning and decision architecture that is trained on vast data and adaptable to a wide range of applications \cite{Bommasani&etal:22}. Some of the defining features of foundation model include the following:
 \bitem
 \item {\bf Architectural partition of pre-training and the adaptation:} The foundation model approach to AI partitions the complex and costly process of {\em foundation model pre-training}---training deep neural networks with billions of parameters and vast training data---from {\em foundation model adaptations} through fine-tuning for specific applications. Such ``division of labor'' is essential to market general AI technologies to a broad range of applications.
 \item {\bf Self-supervised learning of essential features:} For a foundation model to be effective for a wide range of applications means that the foundation model  must capture essential features in vast data, making self-supervised learning essential with unlabeled training data.
 \item {\bf Generative AI:} Foundation models are generative, capable of producing artificial samples not in the training data but with the same characteristics. The generative capability of a foundation model is crucial for generalization and data augmentation.
 \eitem
Most successful AI foundation models are based on large language models (LLMs), pre-trained with vast text, image, and video data. The transformer architecture and attention mechanisms play a critical role in these foundation models to capture naturally highly complex linguistic structures without relying on grammatical rules or stylistic constraints. A key feature of LLM is its ability to model long-range (passage) linguistic dependencies in language.

The adoption of the AI foundation model principle to power systems is well recognized and emerging but still limited. See, \eg
\cite{Majumder&etal:24Joule,Hamann:24Joule,DOE24_rpt} and references therein. This work is the first attempt to develop an AI foundation model for synchro-waveform data generated by WMUs.  Key challenges for data produced by the grid arise from the fundamental differences between grid-generated data and language texts; the former follows the physical laws of Ohms, Kirchhoff, and Maxwell, and the latter the linguistic structures and conventions. The power system signals are sinusoidal; they are generated {\em causally} through a physical process. Typically, they do not have the long-range dependencies as in linguistic passages. It is not clear that the principle of LLM-based foundation model is suitable to power system  applications involving measurements of physical (rather than linquistic) attributes.

\subsection{Innovation Representation Model}
 A generic foundation model structure is an autoencoder that maps the foundation model input to a latent (feature) process by the encoder, and the foundation model input is approximated by the decoder to ensure that the latent feature is, in some sense,  a faithful representation of the input. Typically, the latent process is simpler and has lower dimensionality. For the LLM-based foundation model, the autoencoders are various types of transformers with attention mechanisms.

\begin{figure}[htbp]
    \centering
    \includegraphics[scale=0.5]{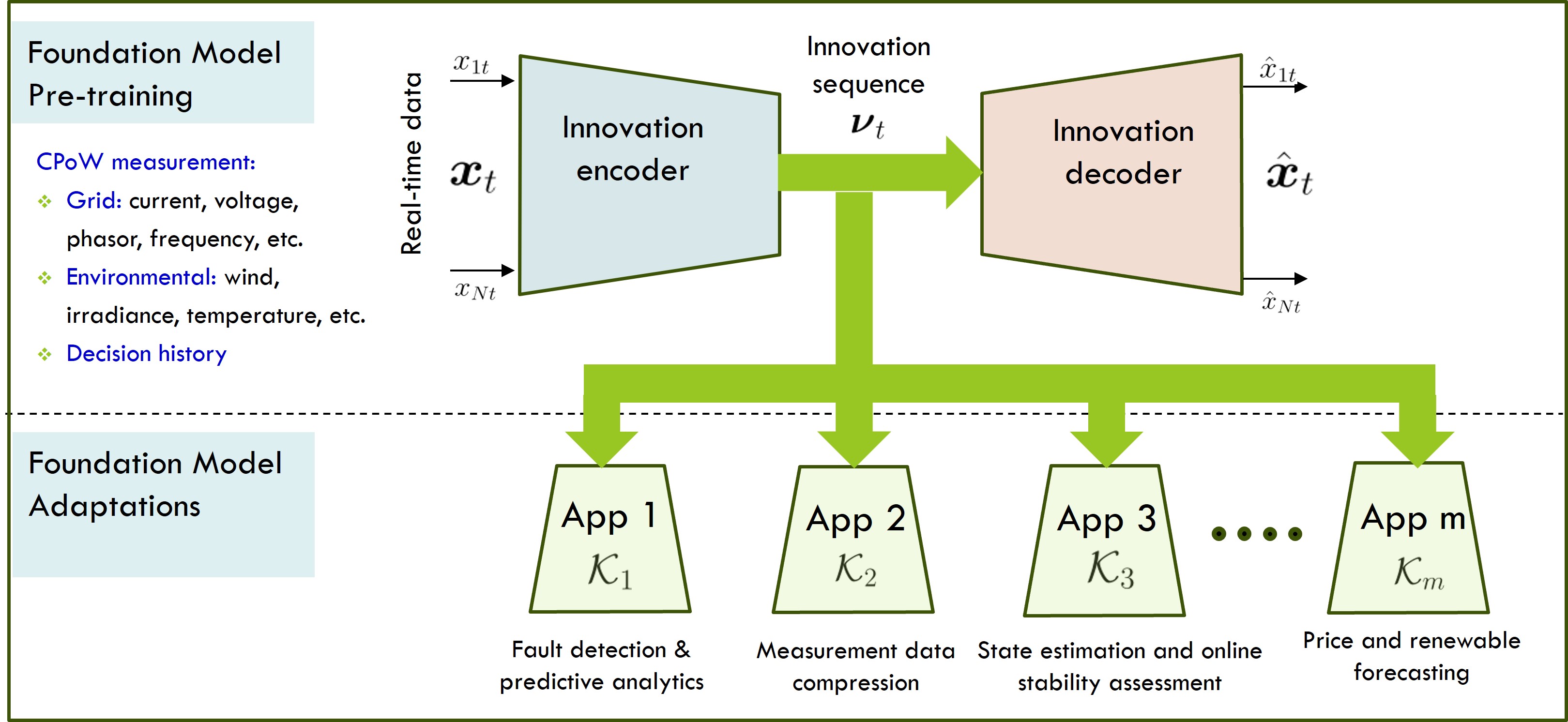}
    \caption{The foundation model pre-training and adaptations with innovation representation model (IRM).}
    \label{fig:FM}
\end{figure}

We propose an innovation-based foundation model for time series from power systems for monitoring and control, having all the general characteristics of foundation model but augmenting the linguistic-based LLM autoencoder with a time series-based {\em Innovation Representation Model (IRM),} denoted by $(\Gc_\theta,\Hc_\eta)$, as shown in Fig.~\ref{fig:FM}.  While LMM-based foundation may be important for broad power system applications, here we focus on power system monitoring and real-time decision-making using power system measurements, where the porposed IRM-based foundation model can be more effective.

The upper part of Fig.~\ref{fig:FM} corresponds to {\em Foundation Model Pre-training} with grid-generated WMU measurement data as well as possible environmental data and decision history.
To enforce the causality of the grid-generated data, $\Gc_\theta$ and $\Hc_\eta$ are convolutional neural networks with parameters $(\theta,\eta)$. The innovation autoencoder $(\Hc_\theta,\Gc_\eta)$ is trained with WMU time series by enforcing the latent process $(\nubf_t)$ to be an {\em innovation process---an independent and identical uniformly distributed (IID) sequence---}of the input $(\xbf_t)$ while the output of the autoencoder $(\hat{\xbf}_t)$ matches with the input either strongly in the mean-squared sense or weakly in the sense of matching distributions.   The pre-training of the foundation model  is discussed in more detail in Sec.~\ref{sec:pretrain}. After pre-training, the neural network parameters are denoted by $(\theta^*,\eta^*)$.

What makes the innovation-based foundation model  ``mathematically correct'' for WMU data is that, when the foundation model is ideally trained, the latent process $(\nubf_t)$ represents {\em sufficient statistics} for the statistical model of $(\xbf_t)$, which implies that any optimal decision based on $(\xbf_t)$ can be implemented without loss using $(\nubf_t)$. Beyond being lossless, IRM also greatly simplifies computation.
While defining the probability structure of the input process, $(\xbf_t)$ is infinite-dimensional that has to account for all possible unknown temporal dependencies, the innovation $(\nubf_t)$ is IID with the same one-dimensional marginal distribution that can be set to be uniform, Gaussian, or other convenient distributions. For example, the uniform distribution has a bounded domain suitable for training, whereas the Gaussian distribution has convenient likelihood expressions, which facilitate effective learning processes.

The lower part of Fig.~\ref{fig:FM} is for {\em Foundation Model Adaptations} for different real-time operation tasks, such as anomaly/novelty detection (Sec. \ref{sec:anomaly}), over-current protection (Sec. \ref{sec:overcurrent}), WMU measurement compression (Sec. \ref{sec:compression}), and probabilistic forecasting \cite{Wang&Tong&Zhao:24arxiv}. For example, to perform fault detection in real-time, current/voltage measurements are passed through the pre-trained innovation encoder $\Gc_{\theta^*}$ to extract innovation sequence $\nubf_t)$ as sufficient statistics, from which the foundation model adaptation module $\Kc_1$ makes optimal detection using, \eg the detector derived in Sec.~\ref{sec:anomaly}.

\subsection{Learning IRM in Foundation Model Pre-training} \label{sec:pretrain}
 IRM was introduced by Norbert Wiener, Gopinath Kallianpur, and Murray Rosenblatt in their seminal work \cite{Wiener:58Book,Masani:66BAMS,Rosenblatt:59,Rosenblatt:09} and more recent work in high-dimensional statistics \cite{Wu:05PNAS,Wu:11}.
Under the Wiener-Kallianpur (strong) innovation representation, there exists a causal autoencoder $(\Gc,\Hc)$ that the encoder $\Gc$ maps causally the input $(x_t)$ to an IID uniformly distributed {\em innovation sequence} $(\nu_t)$, and the causal decoder $\Hc$ that maps $(\nu_t)$ to $(\hat{x}_t \stackrel{\mbox{\tiny a.s.}}{=} x_t)$.  Under the Rosenblatt (weak) innovation representation, the autoencoder input and output are matched in distribution $(\hat{x}_t \stackrel{\mbox{\tiny D}}{=} x_t)$.

\begin{figure}[h]
    \centering
    \includegraphics[width=\linewidth]{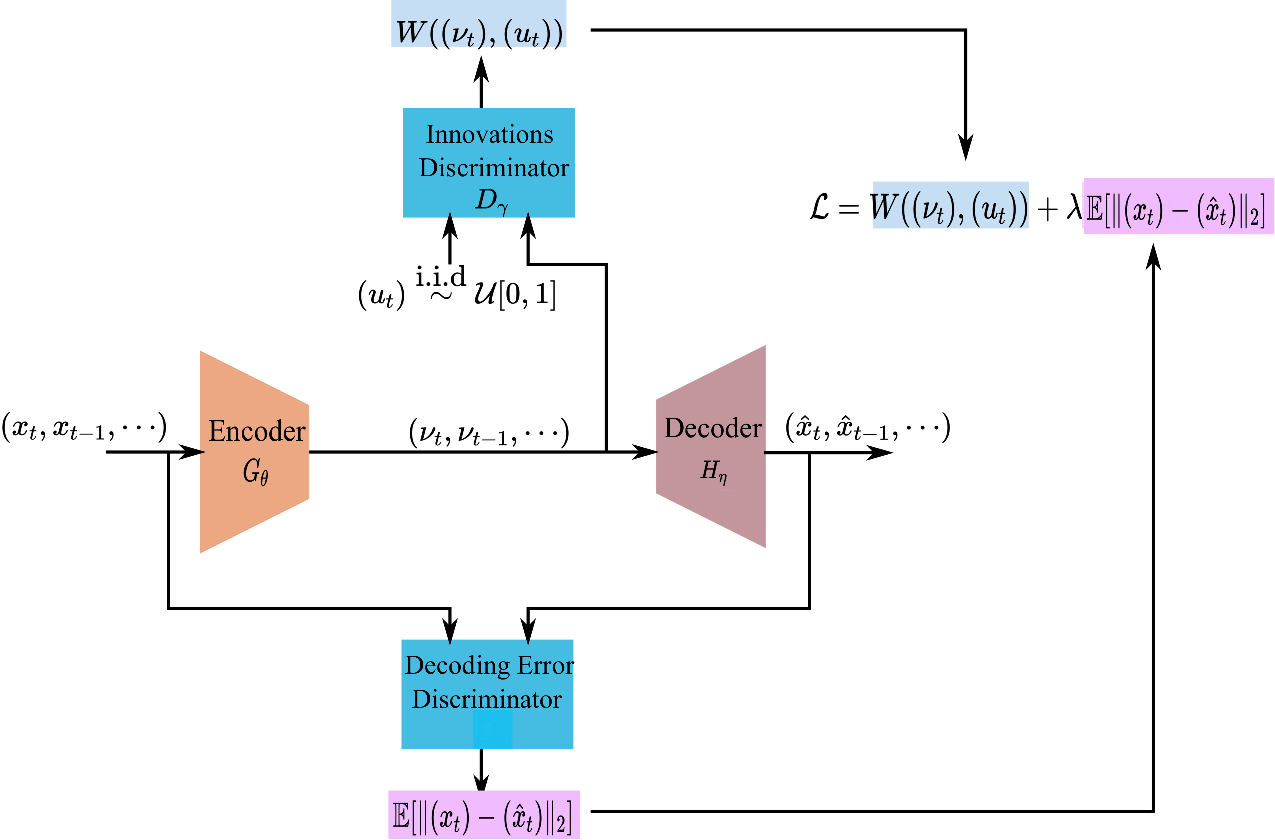}
    \caption{A Machine Learning Architecture for Innovation Representation Model.}
    \label{fig:SIAE}
\end{figure}

Fig.~\ref{fig:SIAE} shows the schematic pre-training of the IRM, first proposed in \cite{Wang&Tong:21JMLR}. The innovation autoencoder is defined by a causal convolution neural network (CNN) encoder $\Gc_\theta$ with parameter $\theta$ and causal CNN decoder $\Hc_\eta$ with parameter $\eta$. These parameters are optimized to match the autoencoder output $\hat{x}_t$ with its input $x_t$ (in the mean-squared sense) and impose the constraint that the latent process $(v_t)$ is IID.

Specifically, the objective function that generates stochastic gradient descents combines the  Wasserstein distance error produced by the {\em innovation discriminator} $D_{\gamma}$ and the mean-squared decoding error between the input $(x_t)$ and output $\hat{x}_t$ of the autoencoder. The joint optimization of the neural network parameters $(\theta,\eta,\gamma)$ is given by
\begin{align}
     L\big(\theta,\eta\big)&:= \max_\gamma \Big\{\tilde{\mbbE}\Big[D_\gamma\big((v^{(\theta)}_t),(u_t)\big)\Big]
     +\lambda \tilde{\mbbE}\Big[ ||(\hat{x}^{(\theta,\eta)}_t)-(X_t)||_2\Big]\Big\}
  \label{eq:loss}\\
  (\theta^*,\eta^*)&:=\arg\min_{\theta,\eta} L\big((X_t),\theta,\eta\big),
\end{align}
where $\tilde{\mbbE}$ is the sample average operator. Note that the $l_2$ norm penalty on the decoding error used in (\ref{eq:loss}) aims to achieve a (strong) mean-squared error reconstruction of the input by the decoder in the Wiener-Kallanpur representation. For the Rosenblatt relaxation of the innovation representation is used, the $l_2$ norm $\mbbE\big[||(\hat{X}_t)-(X_t)||_2\big]$ is replaced by the Wasserstein distance measure $W\big((\hat{X}_t),(X_t)\big)$.

The above optimization can be realized using standard machine learning library \cite{paszke2019pytorchimperativestylehighperformance,tensorflow2015-whitepaper}.  In practice, pre-training operates on finite-dimensional inputs and segments of random processes, with finite neural network architectures. In \cite{Wang&Tong:21JMLR}, the authors established structural convergence, demonstrating that the finite-dimensional implementation converges to the theoretical infinite-dimensional innovation representation as the architecture's complexity increases.

\section{Adaptation for Power System Anomaly Detection}
\label{sec:anomaly}

We present the foundation model adaptation for anomaly detection by formulating a prototype hypothesis testing model involving stochastic WMU measurements $\xbf=(x_t)$:
\beq \label{eq:H0H1}
\Hc_0: \xbf \sim P_0~~\mbox{vs.}~~\Hc_1: \xbf \sim Q  \in \Qmsc,
\eeq
where $P_0$ is the probability model for WMU time series $\xbf$ under {\em normal operating conditions}, and $\Qmsc$ includes all abnormal probability distributions.

The above model (\ref{eq:H0H1}) can be extended for {\em novelty detection,} where the objective is to catalog encountered abnormalities in an anomaly dictionary and detect whether the WMU measurements represent a new type of anomaly not encountered before. This generalization can be formulated as:
\beq \label{eq:H00H1new}
\Hc_0: \xbf \sim P \in \Pmsc~~\mbox{vs.}~~\Hc_1: \xbf \sim Q  \in \Qmsc,
\eeq
where $\Pmsc = {P_0, \dots, P_K}$ is the dictionary of WMU measurement classes previously identified and validated either as known anomalies or as data under normal operating conditions.

In practice, neither $P_0$ nor $\Qmsc$ is known, and the hypothesis testing formulations (\ref{eq:H0H1},\ref{eq:H00H1new}) are ill-defined.  It is the IRM that transforms the ill-defined problem to a well-defined classic hypothesis testing problem.
We assume that large data sets exist for measurements under normal operating conditions, enabling the pre-training of the IRM of the foundation model. Such an assumption is reasonable in practice and necessary for all AI foundation models. With a pre-trained autoencoder, the IRM-based foundation model  transforms (\ref{eq:H0H1}) to a challenging but tractable Goodness-of-Fit test problem:
\beq \label{eq:H0H1_1}
\Hc'_0: \nubf \sim P'_0~~\mbox{vs.}~~\Hc'_1: \nubf \sim Q  \in \Qmsc'.
\eeq
where $P'_0$ corresponds to the IID-uniformly distributed innovation sequence (under normal operating conditions), and $\Qmsc'$ is the collection of probability models for which the latent process of the foundation model encoder is not IID-uniform.

\subsection{Related Work}
The hypothesis testing model in (\ref{eq:H0H1}) falls within the category of nonparametric composite hypothesis testing \cite{Sheskin:07book}, where classical statistical approaches typically assume a known probability model $P$ under $\Hc_0$. Standard techniques, such as the Kolmogorov-Smirnov test and related methods \cite{Sheskin:07book}, are well-suited for offline detection involving non-sequential data under $\Hc_0$. However, these approaches cannot apply to cases when probability models are unknown or when sequential data are involved.

Modern machine learning approaches alleviate the assumption of known probability models under both $\Hc_0$ and $\Hc_1$. Despite this, anomaly and novelty detection in time series remains challenging due to the infinite-dimensional nature of nonparametric time series probability structures. Standard methods address this by approximating the problem in a finite-dimensional space, often segmenting the time series into finite-duration samples (see, e.g., \cite{Ma&Perkins:03IJCNN,Dasgupta&Forrest:1995ICIS,Gardner&etal:2006JMLR}). However, such segmentations often neglect temporal dependencies, which are critical for accurate detection. Large segments can mitigate this issue but result in long detection delays and prohibitive computation costs.

When the probability model under $\Hc_0$ is unknown, state-of-the-art detection techniques can be grouped into three broad categories:
\ben
\item {\bf Classifier-based approaches:} Methods such as one-class support vector machines (SVM) \cite{Scholkopf:99NIPS} and their variants \cite{Khan&Madden:04,Bergmann&etal:19,Gong&etal:19} implicitly assume that the support of the anomalous data distribution is largely disjoint from the anomaly-free distribution. While these techniques are effective in some cases, they perform poorly when the anomaly and anomaly-free models have overlapping support domains.
\item {\bf Simulations with hypothetical anomalies:} These methods simulate samples under $\Hc_1$ using hypothetical distributions that are some distance from $P_0$ \cite{Lee&etal:18ICLR,Hendrycks&etal:18ICLR,Ren&etal:19ICLR}. However, capturing all possible alternative distributions is inherently challenging, resulting in performance gaps caused by anomalies not covered by simulations.

\item {\bf Feature-domain confidence scores:} Many recent machine learning solutions operate in a transformed feature domain, using confidence scores on a learned anomaly-free model to detect anomalies \cite{Hendrycks&Gimpel:17ICLR,Lakshminarayanan&Pritzel&Blundell:17NIPS,Liang&Li&Srikant:18ICLR,Lee&etal:18NIPS}. As shown in \cite{Lan&Dinh:21}, even with a perfect density estimate, these methods may struggle to achieve high detection performance. Autoencoder-based techniques are particularly relevant here \cite{Schlegl&Seebock:19,Bergmann&etal:19,Gong&etal:19,Dinh&Krueger&Bengio:2015,Brakel&Bengio:17}.
\een
The advent of LLM with a transformer architecture and attention mechanisms has shown promise in classification and time series forecasting tasks \cite{Zhou&etal:21AAAI,Zeng&etal:23IAAI}. However, the applicability of transformer-based architectures for anomaly detection in power systems is not well understood. Power system data often lack the long-range dependencies characteristic of natural language, making transformer models challenging to train effectively  \cite{Wang&Tong&Zhao:24arxiv}.

\subsection{Foundation model Adaptation for Anomaly Detection}
With pre-trained IRM, the foundation model adaptation for anomaly detection is to test anomalies in the innovation sequence $(\nubf_t)$ from the innovation encoder output based on the hypothesis testing model (\ref{eq:H0H1_1}). With the null hypothesis being IID-uniform, the original anomaly detection problem (\ref{eq:H0H1}) is transformed by the IRM to the classical Goodness-of-Fit problem, for which considerable literature exists \cite{GOFbook}.

The Neyman's Smooth Test (NST) was first proposed in 1937 \cite{Neyman:37SAJ} for testing IID-uniform samples against the alternative distribution for non-IID uniform samples as in (\ref{eq:H0H1_1}). In the contemporary machine learning language,  NST can be viewed as a kernel-based classifier using the orthogonal Legendre polynomial  $\{\pi_k\}$ as a kernel representation of probability distributions under $\Hc_1$.

\begin{figure}[h]
    \centering
    \fontsize{10pt}{3pt}
    \includegraphics[scale=0.6]{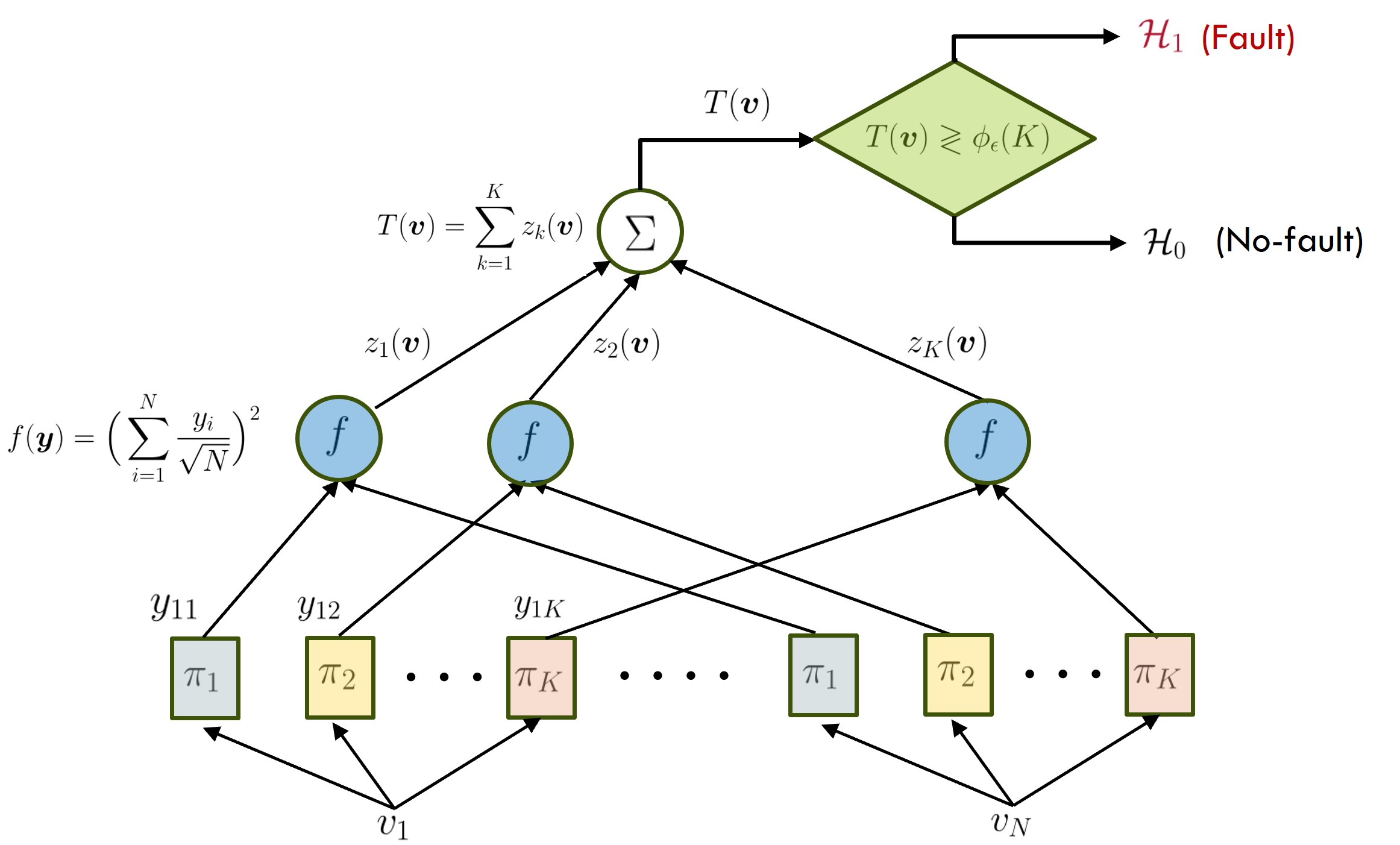}~~~~
\vspace{0.5em}
\caption{Neyman's Smooth Test (NST).}
\label{fig:NST}
\end{figure}

The schematic of NST is shown in Fig.~\ref{fig:NST} where the innovation input {$(v_t)$} are passed through Legendre polynomial kernels at the input layer. The output of the input layer is the sum of independent random variables at the Central-Limit Theorem scaling, resulting in a limiting Gaussian distributed $\sum_i y_i/\sqrt{N}$, and $\chi^2$ distributed $T(\vbf)$ with $K$-degree of freedom, asymptotically. The testing is a threshold decision based on
\begin{align}
\sum_{i=1}^K\left(N^{-1/2}\sum_{t=1}^N\pi_i(v_t)\right)^2:=T(\vbf)\tmb{\Hc_1}{\gtrless}{\Hc_0} \phi_\varepsilon(K),
    \label{Eq:decision}
\end{align}
where $\phi_\varepsilon(K)$ is the $(1-\varepsilon)$-quantile of a Chi-square distribution with $K$ degrees of freedom.  Parameter $\varepsilon$ can be chosen according to the desired false positive rate.

NST was shown to be the locally most powerful test for uniformity \cite{Neyman:37SAJ}, outperforming all other possible detection techniques in a neighborhood around the null hypothesis $\Hc_0$.  In \cite{Rayner&Best:90}, Rayner and Best declared: ``don't use those [other] methods---use a smooth test!''

NST was originally designed for fixed-sample size testing. When testing time series where measurements arrive sequentially, a sequential test can result in a considerable speed-up in detection time. Here, we propose a sequential NST (SNST)  to reduce detection errors and decision time. SNST is a data-driven approach that makes testing decisions using the latent innovation sequence of the innovation encoder. It is designed to achieve short detection delay by implementing a sequential version of NST, following the framework of change detection. When distribution statistic shows that the distribution of $(\nu_t)$ exhibits significant deviation from the IID-uniform characteristics, the procedure terminates quickly by declaring fault. When $(\nu_t)$ remains close to IID-uniform, no alert is made as more data samples are collected to enhance detection accuracy.

\begin{algorithm}[h]
   \caption{A doubling search implementation of a $K$-th order Neyman's smooth test at time $t^*$.}
   \label{alg:NS_DST}
\begin{algorithmic}
   \State {\bfseries Input:} innovations data $(v_t)$, order of Legendre polynomials $K$, minimum separation $\lambda$, target false positive rate $\varepsilon$, and a scaling constant $C$.
   \State {Compute threshold $\phi_\varepsilon(K) :=\inf\{\phi:\Pr[Q>\phi]\leq\epsilon,Q\sim\chi^2(K)\}$}
   \State Output $=\mathcal{H}_0$.
   \For{$i=1,2,\cdots,\log_2 \lambda$}
        \State $N = 2^iC$
        \State Draw $N$ samples of $(v_{t^*},v_{t^*+1},\cdots,v_{t^*+N-1})$
        \If {$N^{-1/2}\sum_{i=1}^K\sum_{t=t^*}^{t^*+N-1}\pi_i(v_t)>\phi_\varepsilon(K)$}
            \State Output $=\mathcal{H}_1$
            \State Break
        \EndIf
   \EndFor
    \State Report Output to the executor (circuit breaker).
\end{algorithmic}
\end{algorithm}

SNST implements the above sequential procedure using the so-called doubling search technique \cite{OufkirEtalNeurips2021}. At each iteration, SNST executes NST with a fixed-sized batch of samples. If the output of the detecting technique is $\Hc_0$, the batch size will be doubled, and the detecting technique will be re-run in the next iteration. An upper bound on the number of iterations is often set to avoid infinite detection delay. The detailed algorithm of the doubling search trick implementation of NST is presented in Algorithm~\ref{alg:NS_DST}.
Once NST has made its detection decision, its decision is passed to the circuit breaker to carry out its protection action.

\section{Adaptation for Over-Current Protection}
\label{sec:overcurrent}

Traditional power system protections are based on a deterministic power system model. For short-circuit faults, the standard practice is to pre-set the threshold (the pickup current) such that current flows exceeding the threshold trigger protection relay actions.

\begin{figure}[h]
    \centering
    \includegraphics[scale=0.3]{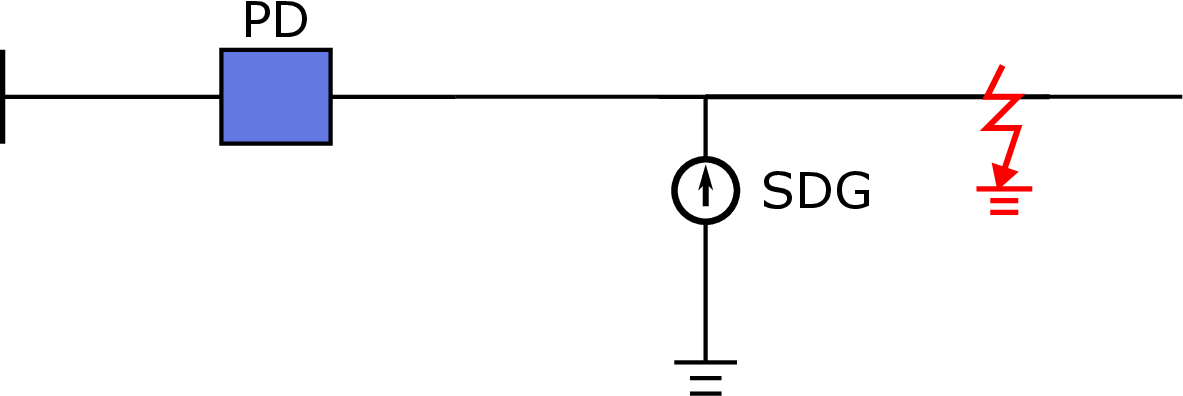}~~~~
 \includegraphics[scale=0.3]{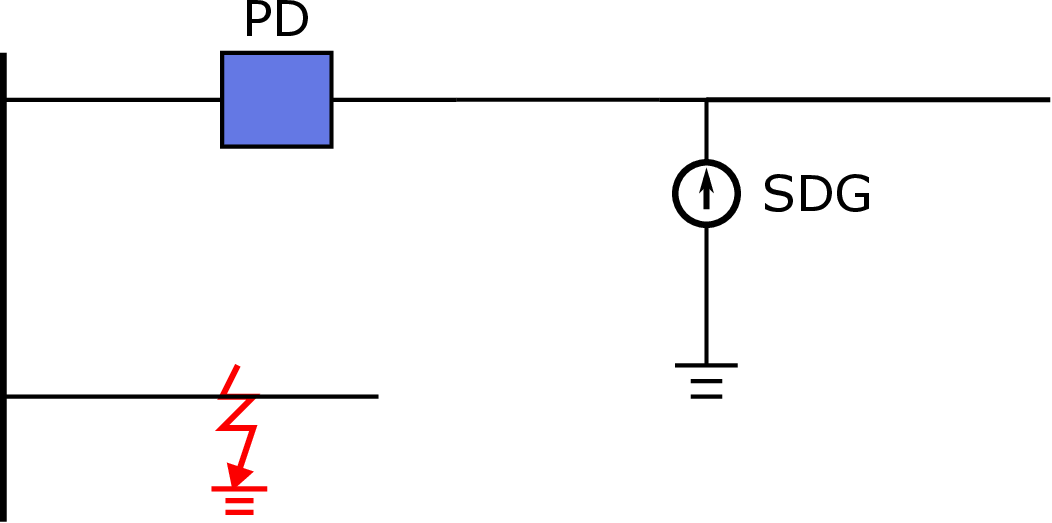}
\vspace{0.5em}
\caption{Two types of erroneous protection actions caused by stochastic distributed generation (SDG). Left: Protection blinding. Right: Sympathetic tripping.}
\label{fig:tripping}
\end{figure}

The significant presence of stochastic distributed generation (SDG) and IBRs causes two types of erroneous relay actions shown in Fig.~\ref{fig:tripping}. One is the so-called {\em protection blinding} shown on the left panel, where SDG offsets the high current flow through the protection device (PD), causing a Type II  (miss-detection) error in the protection decision. The other type is the {\em sympathetic tripping} shown on the right panel, where the high current flow from the SDG causes the Type I (false alarm) detection error that triggers protection action when it should not be activated.

\subsection{Related Work}
Power system protection under significant SDG is an active area of research because of the increasing integration of power electronics-based resources. In \cite{TelukuntaEtal17:CSEEPES}, a detailed characterization and classification of distribution system protection under bulk renewable energy sources is provided. Classical techniques typically compare the measurements with pre-fixed thresholds \cite{Anderson99}. Other methods include installing fault current limiter \cite{Elkhatam&Sidhu08TPD,Ibrahim17IJEPES} to mitigate the effect of SDG or treating pickup current as a decision variable in an optimization program, taking into account network configurations \cite{ChattopadhyaySachdevSidhu96:TPD,LiuEtal:16CSEEPES}.

More sophisticated data-driven solutions have been proposed. Methods belonging to this category set current threshold settings based on either estimated SDG levels \cite{WanLiWong10:TIA,PapaspiliotopoulosEtal:17TPD,Rezaei19EEEIC}.  Relay settings can also be computed from estimated system parameters of a Thevenin equivalent system \cite{ShenEtal:17TPD}, steady-state fault current \cite{MaEtal:12EPES}, and state estimation
\cite{Meliopoulos&etal:17TPS}. Other data-driven methods are developed for networked sensors as a multi-agent system \cite{LiuEtal:17TPD}, assuming a network supporting communications among agents. The method proposed in \cite{MahatEtal:11TSG} adjusts the over-current relay setting based on current measurements from relays downstream.

\subsection{Adaptation for Data-driven Over-current Protection}
We propose an IRM-based foundation model approach, adapting to the sequential anomaly detection approach described in Sec.~\ref{sec:anomaly} for data-driven over-current protection in the presence of significant SDG, where WMU measurements are used for detecting over-current faults.

\begin{figure}[h]
    \centering
    \includegraphics[scale=0.6]{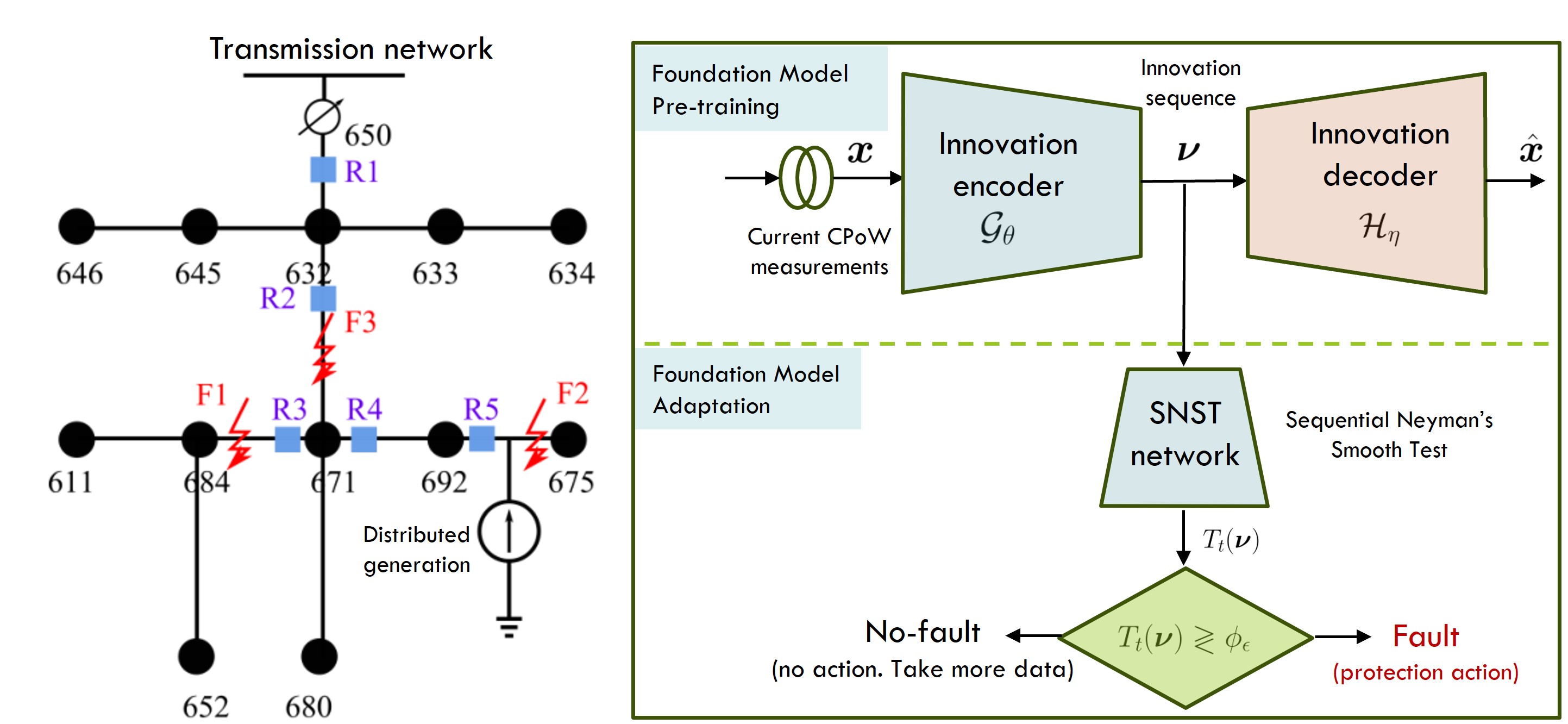}
\vspace{0.5em}
\caption{Schematic of the foundation model adaptation for over-current protection. Left panel: the IEEE 13 bus distribution system where protection relays R1 to R5 are labeled by the squares. Right panel: the implementation of SNST for over-current fault detection.}
\label{fig:Protection}
\end{figure}

Fig.~\ref{fig:Protection} shows a schematic for the foundation model adaptation for over-current protection. The left panel shows the IEEE 13 bus distribution system where protection relays R1-R5 are deployed. Short-circuit faults may occur at various locations, as illustrated by F1 to F3. With SDG injecting stochastic renewable power between bus 692 and 675, it will impact the primary relay action of R5, secondary relay R4, and possibly others.

The proposed foundation model adaption solution for over-current protection is shown in the right panel of Fig.~\ref{fig:Protection} that implements the sequential anomaly detection algorithm SNST discussed in Sec.~\ref{sec:anomaly}. With the pre-trained foundation model, the innovation autoencoder installed at each relay takes the WMU (current) measurements $\xbf:=(x_t)$, extracts the innovation sequence $\nubf:=(\nu_t)$, and passes $\nubf:=(\nu_t)$ through the neural network that generates Neyman's smooth test statistic. At time $t$, the current and past $N$ innovations samples of $\nubf$ are the input of the
SNST adaptation network whose output  $T_t(\nubf)$ is the test statistic at time $t$ against threshold $\phi_\epsilon$ designed to control the Type I error rate at $\epsilon$. If $T_t(\nubf)>\phi_\epsilon$,
the over-current fault is flagged. Otherwise, no relay action is taken, and more data samples are collected.

\subsection{Baseline Comparisons}
{We compared the proposed foundation model adaptation for over-current protection with two  protection baselines: (i) the  ``conventional" over-current relay protection with a predetermined pickup current and the inverse time function defined by the IEEE 37112 standard \cite{IEEEStd37112-2018}, and (ii) AOCR---an adaptive method proposed in \cite{JainEtal19TPD}--with pickup current as a linear function of $10$-second moving-average current and minimum fault current in the protection zone. All three methods required setting detection thresholds. The SNST threshold was obtained analytically for a given false positive rate (FPR). The threshold settings for the conventional and AOCR methods were obtained empirically through simulations to meet the FPR target.}

{For the proposed foundation model in the experiments, the learning of the IRM autoencoder pair and the discriminator in Fig~\ref{fig:SIAE} are convolutional neural networks with three layers. The filter size is set to 20, and the first layer's input dimension is 50. Each layer's output dimensions are 100, 50, and 25. Batch normalization is applied to the input layer. The Adam optimizer is used to get the optimal weights for all neural network structures.  The SNST used in the foundation model adaptation was implemented based on a $4$th-order Neyman's smooth test with the doubling search procedure as described in Algorithm~\ref{alg:NS_DST}. We chose $C=42.5$ and $\lambda=20$. We generated $10,000$ samples under the anomaly-free condition for innovation autoencoder training. We adopted the Adam optimizer with $lr = 1e-5$ to train the innovation autoencoder. The threshold of the conventional method and AOCR are chosen manually to cap the FPR at $0.05$. The inverse-time characteristics of these two over-current relay-based methods are chosen according to the moderately inverse characteristics in IEEE standard \cite{IEEEStd37112-2018}.}

We evaluated the performance of each protection scheme by the True Positive Rate (TPR), the False Positive Rate (FPR), and the detection delay. For each fault scenario, we conducted $1,000$ independent runs of the same system, driving the SDG current profile with $1,000$ different trajectories learned from the EPFL dataset.

The trajectories were learned through a weaker version of innovation autoencoder (WIAE) with the $l_2$ distance between $(X_t)$ and $(\hat{X}_t)$ in the learning objective replaced by the Wasserstein distance. WIAE was trained by the EPFL current profile, and SDG current trajectories from the same distribution as the EPFL current profile can then be drawn as the output sequence $(\hat{X}_t)$ of the decoder.
The detailed implementation of WIAE can be found on Github\footnote{\url{https://github.com/Lambelle/WIAE}}. All metrics were calculated through  $1,000$ Monte Carlo runs.

We evaluated the performance of three candidate solutions using the IEEE 13-bus distribution network shown in Fig.~\ref{fig:Protection} (left), where multiple fault scenarios were considered in our study. To account for randomness in the measurements introduced by renewable energy resources, we added a $1$ MW SDG controlled by a real-world current profile collected at the micro-grid located in EPFL between node $692$ and $675$. The SDG was driven by EPFL's current profile at the battery connection\footnote{Data can be found here: \url{https://github.com/DESL-EPFL/Point-on-wave-Data-of-EPFL-campus-Distribution-Network}}, and current measurement samples were collected at $50$ KHz sampling frequency. The entire system is simulated through MATLAB/Simulink.

\subsection{Numerical Results}
When a single-phase short circuit fault happens at F1, R3 is the primary relay for F1 and R2 as a backup. Sympathetic tripping at R4 may occur under F1 because the current feed from SDG downstream could increase significantly. Similarly,  R4 is also subject to sympathetic tripping under F3 due to current flow from downstream DG. In this case, the primary relay is R2, and the backup is R1. On the other hand, the primary relay R5 for fault scenario F2 faces protection blinding because of DG's support for current and voltage at R5 during fault at F2.

Table~\ref{tab:13bus results} shows the test results. For each method deployed at the locations of different relays, their TPR, FPR, and delays were computed from Monte Carlo runs conditioned on no-faults and faults F1, F2, and F3.
\begin{table}[htbp]
    \centering
    \caption{Detection Results for Different Fault Scenarios of IEEE 13-bus System.}
    \begin{tabular}{c|c|c|c|c|c}
    \hline
         Location & Relays & Method & TPR &FPR &Delay(s)  \\
         \hline\hline
        \multirow{9}{*}{F1} &\multirow{3}{*}{\makecell{R3\\(primary)}} &SNST &1 &0.0451 & 0.0017 \\
        \cmidrule{3-6}
        & &AOCR &0.9893&0.0309&0.0175\\
        \cmidrule{3-6}
        & & Conv. &0.9940&0.0402& 0.0102\\
        \cmidrule{2-6}
        & \multirow{3}{*}{\makecell{R2\\(backup)}} & SNST &0.9991 &0.0255& 0.0068\\
        \cmidrule{3-6}
        & &AOCR &0.9894 &0.0097& 0.0236\\
        \cmidrule{3-6}
        & &Conv. &0.9942 &0.0500 &0.0143\\
        \cmidrule{2-6}
        & \multirow{3}{*}{\makecell{R4\\(sympathetic tripping)}} & SNST &\bf 0.0779 &0.0453 &N/A\\
        \cmidrule{3-6}
        & &AOCR &\bf 1.0 &0.0052&0.0301\\
        \cmidrule{3-6}
        & &Conv. &\bf 1.0 &0.0093& 0.0441 \\
        \hline
        \multirow{6}{*}{F2}& \multirow{3}{*}{\makecell{R4\\(backup)}} & SNST &0.9982 &0.0492 &0.0136\\
        \cmidrule{3-6}
        & &AOCR &0.4260 &0.0414&0.0340\\
        \cmidrule{3-6}
        & &Conv. &0.5569 &0.0450&0.0133\\
        \cmidrule{2-6}
         &\multirow{3}{*}{\makecell{R5\\(primary)}} &SNST&\bf 0.9875 &0.0371 &0.0017 \\
        \cmidrule{3-6}
        & &AOCR &\bf 0.7118 &0.0490 &0.0221\\
        \cmidrule{3-6}
        & & Conv. &\bf 0.6425 &0.0415 &0.0053\\
        \hline
        \multirow{6}{*}{F3} &\multirow{3}{*}{\makecell{R1\\(backup)}} &SNST &0.9927 &0.0482 &0.0068\\
        \cmidrule{3-6}
        & &AOCR &0.8912 &0.0409 &0.0234\\
        \cmidrule{3-6}
        & & Conv. &0.4447 &0.0341 & 0.0211\\
        \cmidrule{2-6}
        & \multirow{3}{*}{\makecell{R2\\(primary)}} & SNST &0.9818 &0.0385 &0.0017\\
        \cmidrule{3-6}
        & &AOCR &0.8847 &{ 0.0501}&0.0233\\
        \cmidrule{3-6}
        & &Conv. &0.4487 &{ 0.493}& 0.0187\\
        \cmidrule{2-6}
        & \multirow{3}{*}{\makecell{R4\\(sympathetic tripping)}} & SNST &\bf 0.0593 &0.0257 &N/A\\
        \cmidrule{3-6}
        & &AOCR &\bf 0.9939 &0.0025&0.0311\\
        \cmidrule{3-6}
        & &Conv. &\bf 1 &0.0043&0.0422\\
        \hline
    \end{tabular}
    \label{tab:13bus results}
\end{table}

\noindent\textbf{True Positive Rate Performance:} From the TPR column evaluated under faults F1, F2, and F3, respectively, the primary and backup relays should trip for all faults. We see that SNST achieved the best detection accuracy for the primary and backup relays whenever a relay served as either a primary or a backup under all three faults. For example, under F1, the fault detection rate is $100\%$ at the primary relay R3 and $99.91\%$ at the backup relay R2. Similar performance were observed under F2 and F3. SNST did have sympathetic tripping due to false positive detection at Relay R4  $7.8\%$ of the time due to the presence of SDG.
{SNST improved TPR in primary relay by 5.7\%, 106.9 \%, and 67.3\% under F1, F2 and F3, respectively.}

Both AOCR and the conventional did not perform well at the
primary relay {under F2 and F3}, with detection rates under $90\%$. At the backup relays, AOCR and the conventional performed well under F1 but insufficient under F2 and F3. Sympathetic tripping occurred $100\%$ of the time under F1 and F3.

\vspace{0.5em}
\noindent\textbf{False Positive Rate of Detection:}
All methods achieved FPR smaller than $5\%$, since we computed their corresponding decision threshold by setting their empirical FPRs to be $0.05$, analytically or empirically. In particular, for SNST, the decision threshold is computed analytically, and its FPR under all cases is smaller than $5\%$, showing that the innovation autoencoder transformer current measurements to IID uniform samples effectively.

\vspace{0.5em}
\noindent\textbf{Detection Delay Performance:}
The delay column shows how quickly the detection decisions were rendered by the three methods.
{Detection delay is defined as the latency between fault time and fault detection time.}
SNST significantly outperformed both AOCR and the conventional, reducing the detection time by 90\% at primary relays. At backup relays, SNST performed better for most cases except under F1 (worse than AOCR) and under F2 (comparable with the conventional).

\begin{figure}[h]
    \centering
    \includegraphics[width=\linewidth]{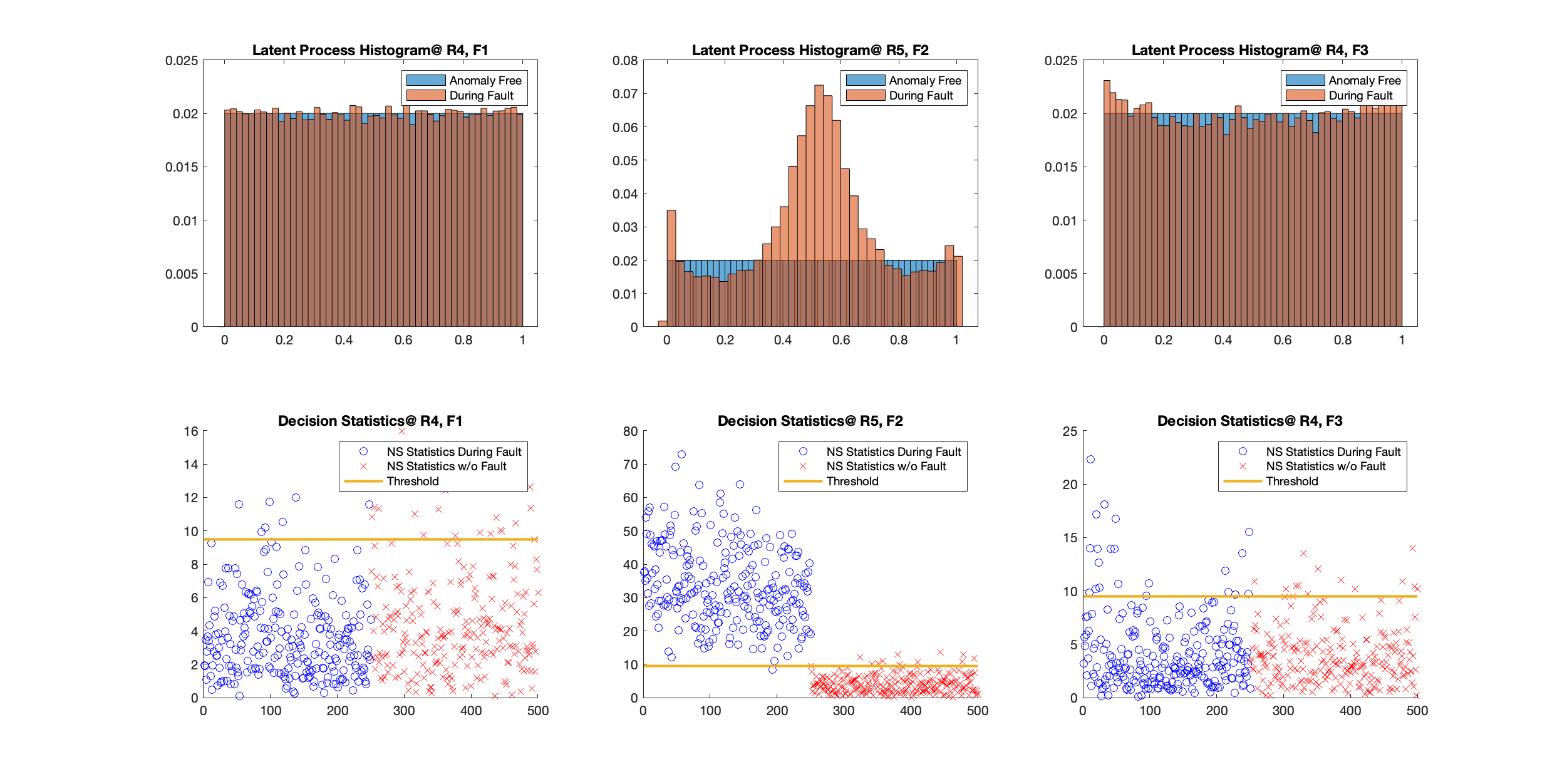}\\
    \caption{Top row: Histogram of the latent sequence generated by the innovation autoencoder. Bottom row: the scatter plots of decision statistics.}
    \label{fig:IEEE_inn}
\end{figure}

To gain insights into the test results, we plotted the decision statistics of each method with 250 Monte Carlo runs each for both anomaly and anomaly-free cases, marked in circles and crosses, respectively.

\vspace{0.5em}
\noindent\textbf{Insights from test statistics for the SNST detector:} For the proposed SNST detection, we plotted the histogram of the innovation sequence associated with the current WMU measurements under three faults in Fig.~\ref{fig:IEEE_inn}. The top row of the figure shows the histograms of the latent processes under three faults. The bottom row shows the scatter plots of SNST's decision statistics.

The leftmost panel shows the case of F1 at R4 unrelated to F1. For this case, the test should have been negative had there been no SDG. The top-left and top-right figures show that the histograms of the latent processes were approximately uniform, indicating that the observations were deemed normal. The scatter plots below show that the Neyman smooth test statistics were below the threshold most of the time. Thus, the rates of positive tests were about $7\%$ and $5\%$, respectively.

The middle panel shows the case of F2 at R5. As the primary relay, R5 should trip with low detection delays. The top-middle figure shows that the distributions of innovations under fault and no-fault cases exhibit clear distinctions. The anomaly-free histogram was uniform, whereas the anomalous one showed a significant deviation from the uniform distribution. Consequently, Neyman smooth test statistics for the anomalous case were mostly above the decision threshold, resulting in a TPR higher than $98\%$.

\begin{figure}[h]
    \centering
    \includegraphics[width=\linewidth]{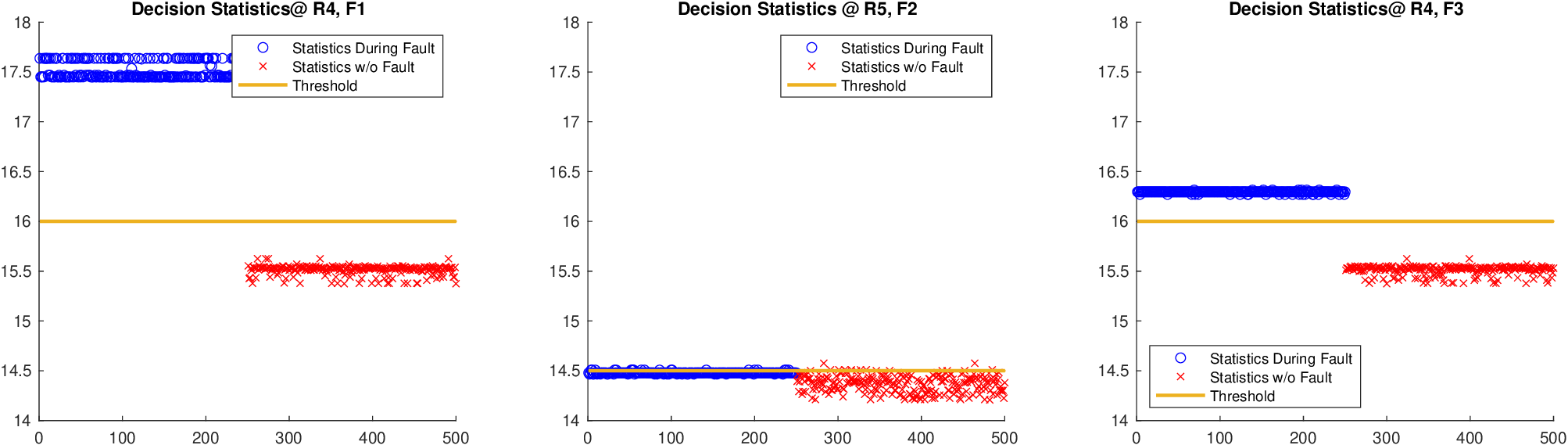}
    \caption{Scatter Plot of Decision Statistics used in the conventional protection method. In other words, the maximum current magnitude.}
    \label{fig:Conv_stats}
\end{figure}

\vspace{0.5em}
\noindent\textbf{Test statistics for the conventional detector:}
For the conventional method, the test statistic was based on the pickup current. We plotted the maximum current magnitude in each decision block with decision threshold $I_{pickup}$, shown in Fig.~\ref{fig:Conv_stats}.

The leftmost panel shows the case of F1 at R4, where the current magnitude showed a slight increase due to the presence of SDG. This challenges the detection accuracy of the conventional method, where the decision is made solely based on the current magnitude.
As seen from the leftmost panel of Fig.~\ref{fig:Conv_stats}, since the magnitude of current during fault at R4 is higher than its no-fault counterpart, R4 trips $100\%$ of the times, leading to the sympathetic tripping phenomenon. A similar pattern is shown under the case F3 at R4, as supported by the rightmost panel.

The middle panel displays the current magnitude under the case of F2 at R5. Due to R5's geographical closeness to the SDG, the no-fault current at R5 exhibits great randomness, and the fault current is decreased compared to the no-SDG counterpart, as supported by the middle panel of Fig.~\ref{fig:Conv_stats}. Hence, achieving high TPRs while maintaining FPRs smaller than $5\%$ is not attainable with a single pre-set threshold.

\begin{figure}[htbp]
    \centering
    \includegraphics[width=\linewidth]{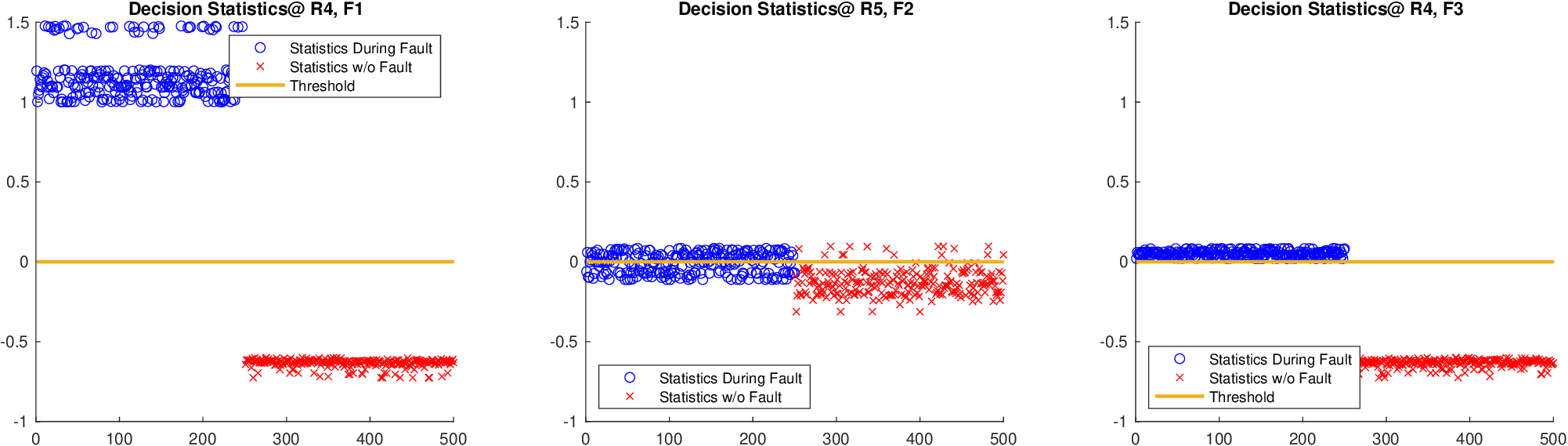}
    \caption{Scatter Plot of Decision Statistics used in the AOCR.}
    \label{fig:AOCR_stats}
\end{figure}

\vspace{0.5em}
\noindent\textbf{Test statistics for the AOCR detector:}
AOCR compared the maximum current magnitude with an adaptively chosen threshold. We plotted the difference between the maximum current magnitude and the adaptively chosen thresholds in each decision block in Fig.~\ref{fig:AOCR_stats}. A difference greater than $0$ implies fault for AOCR.

The decision statistics shown in the leftmost panel of Fig.~\ref{fig:AOCR_stats} were obtained under the case F1 at R4, where SDG induced an increase of fault current at R4, which should not have tripped during fault. Though AOCR computed the decision threshold adaptively, it still compared the pre-fault and post-fault current magnitudes at R4. The statistics showed that the discrepancy between anomaly-free and anomalous cases was smaller using AOCR rather than the conventional method, but the negative impact of the current magnitude increase on decision statistics persisted. For $100\%$ of its time, the current magnitude at R4 exceeded the adaptive threshold computed based on a 10-second moving average of pre-fault current.

The middle panel displays the current magnitude under the case of F2 at R5, where R5's geographical closeness to the SDG introduced significant randomness to the no-fault current. The fault current witnessed a decrease in magnitude compared to the no-SDG counterpart, leading to inaccurate detection results. Though AOCR tried to mitigate the problem by choosing the threshold adaptively, it could not accommodate the high volatility in SGD's output power, resulting in lower detection accuracy.

\section{Adaptation for Synchro-waveform Compression}
\label{sec:compression}
This section presents the foundation model adaptation to compression of synchro-waveform streaming with the goal of facilitating communications by allocating the same or slightly higher communication bandwidth required by PMU data streaming systems. The raw sampling rates of WMU data are one to three orders of magnitude higher than PMU, as shown in Fig.~\ref{fig:events}. To capture some of the IBR-related events at the same bandwidth as required by PMU systems, WMU data streaming needs to have compression ratios of 10-100:1 while maintaining the same level of decompression accuracy.

We propose here a foundation model adaptation for lossy WMU data compression, built on the subband coding idea first proposed in \cite{Wang&Liu&Tong:21TPS}. Instead of compressing raw WMU streaming, we propose compressing and communicating the innovation sequence extracted by the IRM, which is the data feature sufficient for monitoring, control, and real-time decision-making.

\begin{figure}[h]
    \centering
    \includegraphics[scale=0.5]{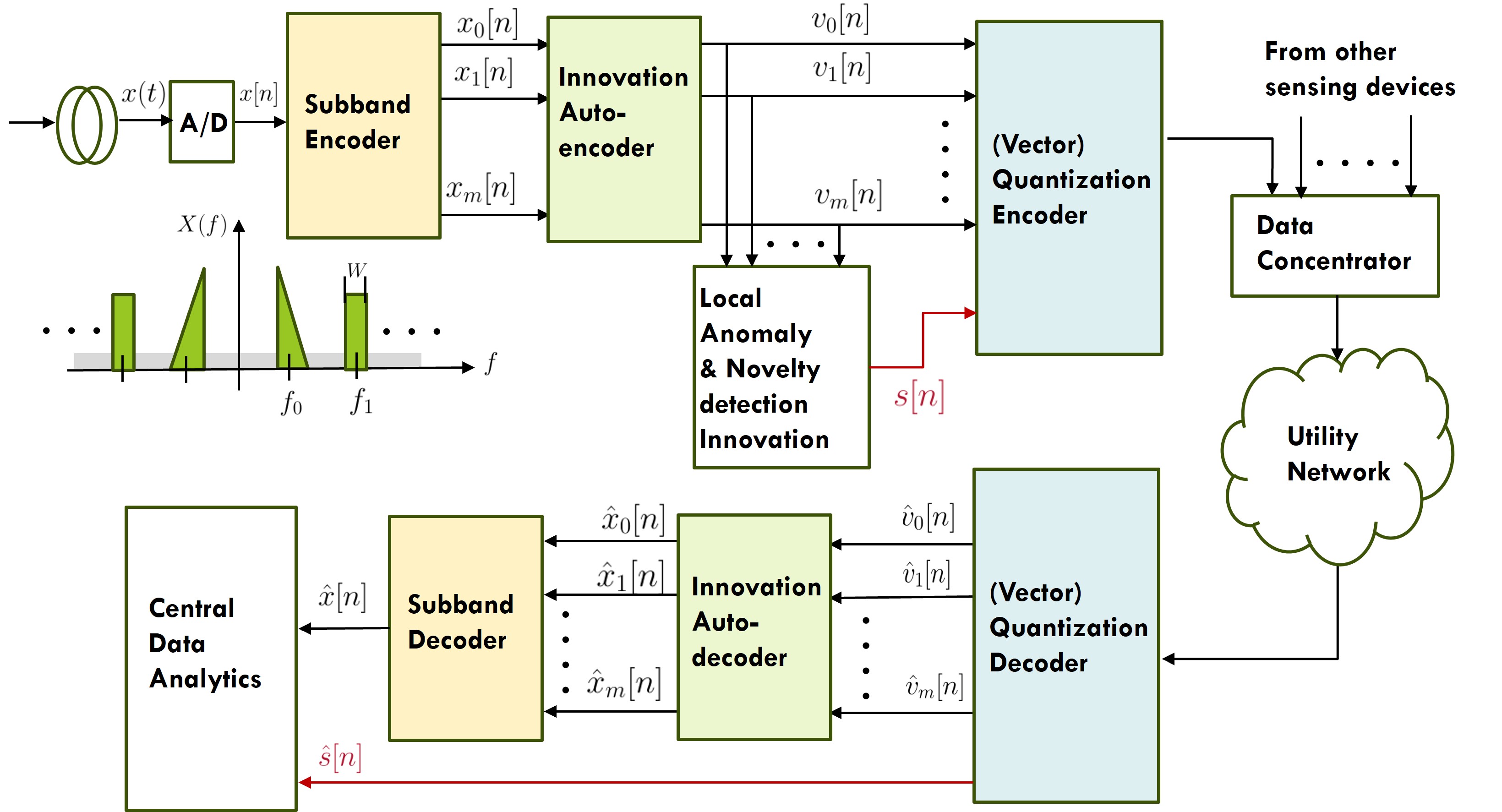}
    \caption{A schematic of the foundation model adaptation for WMU data compression.}
    \label{fig:IBC}
\end{figure}

The schematic of the proposed compression system is shown in Fig.~\ref{fig:IBC} with key components explained next. The upper branch of Fig.~\ref{fig:IBC} is the foundation model adaptation schematic in the intelligent electronic devices (IEDs) where current and voltage WMU data are generated. The lower branch resides in the substations or control centers where monitoring and control decisions are made.
Power system signals are narrow passband signals with special subband spectral structure illustrated in Fig.~\ref{fig:IBC}. Around the fundamental frequency $\pm f_0$ is the fundamental frequency band regulated with frequency deviations within $W_0$. Typically $W_0 < 1$ Hz. Above the fundamental frequency band are harmonic subbands at the multiples of $f_0$, $f_k = k f_0$, that account for higher-order harmonics due to nonlinear devices in the network. There may also be inter-harmonic frequency components represented by the grey area. Higher-order harmonics and inter-harmonics may appear sporadically; they are modeled as on-off random processes.

\paragraph{Subband Coding}
The A/D converter samples the WMU $x(t)$ at the Nyquist rate to obtain a digital continuous point-on-wave sequence $x[n]$. Depending on the spectral bandwidth of $x(t)$, the Nyquist sampling rate is typically in the tens of kHz (or higher) range.
The subband coding scheme proposed in \cite{Wang&Liu&Tong:21TPS} greatly reduces the Nyquist-sampled data rate via the subband decomposition that encodes each harmonic (and inter-harmonic) subband separately. By the {\em passband Nyquist sampling theorem,} a harmonic subband with passband bandwidth $W$ requires only a $2W$ sampling rate. For instance, if the fundamental subband has the bandwidth of $W_0=1$ (Hz), two samples per second or one sample per 30 cycles is sufficient to capture the signal in the fundamental frequency band fully. For an $m$ harmonic-subband spectrum, each with $W$ Hz passband bandwidth, $mW/30$ samples per cycle are sufficient for lossless compression. The subband coding block extracts the baseband components of $X(f)$, down-sampling the Nyquist-sampled data to achieve two to three orders of magnitude data rate reduction. The subband coding technique is well developed for multi-media compression and is embedded in practical audio/video communications data-streaming technologies. The subband encoder proposed in \cite{Wang&Liu&Tong:21TPS} can be easily implemented by off-the-shelf filter banks. See details in \cite{Wang&Liu&Tong:21TPS}. Note the per-cycle data samples of the subband WMU signals $\{(x_i[n])\}$ are baseband representations of the signal spectrum, typically at the rate of less than one sample per cycle under normal operating conditions.

\paragraph{Innovation Autoencoder and Local Data Analytics}
While the subband coding scheme significantly reduces the data rate of the direct Nyquist sampled $(x[n])$, further data rate reductions can be achieved through information-theoretic approaches that offer an optimal tradeoff between the compression rate and distortions of reconstruction. Additional data rate reduction can be achieved through local data analytics that adaptively select harmonic subbands for transmission. For instance, harmonic and inter-harmonic subbands may be inactive under normal operating conditions. During significant network events, signal contents from only a few harmonic subbands may be informative.

The innovation autoencoder of the pre-trained foundation model  discussed in Sec.~\ref{sec:FM} is used to transform subband signals $\{(x_i[n])\}$ to a set of innovations sequences $\{(v_i[n])\}$, each is an IID uniform or Gaussian sequences. For optimal data compression over the typical additive Gaussian noise channel of the utility network with innovation sequence $(v_i[n])$, the optimal bits-allocation to each subband innovation can be determined.
This is discussed as part of the vector quantization encoder in Sec.~\ref{sec:vector_quantization}.

The local data analytics unit detects anomalies and novelties using techniques presented in Sec.~\ref{sec:anomaly}. The local data analytics unit produces the local state sequence $(s_t)$ that can be used to compress data further and provide summary statistics for the control center. Specifically, $(s_t)$ may include the normal/fault state sequence and estimated fault locations when faults are detected locally. Under the normal system state, data from many subbands can be suppressed, and the control center can recall the suppressed signal if necessary.

\paragraph{Vector Quantization Encoder}
\label{sec:vector_quantization}
Vector quantization converts innovations and state sequence $\{(\nu_i[n]), s[n])\}$ to bit-streams to be transmitted over the utility network (after multiplexing with bit streams from other IEDs.)
The quantization schemes developed in \cite{Wang&Liu&Tong:21TPS} can be used. The new feature considered here is an information theoretic approach to compression not included in \cite{Wang&Liu&Tong:21TPS}.

Consider the IID-Gaussian innovation sequences $(\nu_i[n])$, with mean and variance $(\mu_i,\sigma_i^2)$ for the $i$th subband innovations.
Given the acceptable mean-square distortion of the reconstruction $D_i:=\mbbE(\hat{x}_i[n]-x_i[n])^2$, the optimal compression is achieved by allocating distortion $D_i$ to subband $i$ constrained by $D=\sum_i D_i$ through the so-called inverse water-filling procedure \cite{Cover&Thomas:book}.
The Gaussian rate-distortion function gives the minimum transmission rate subject to the distortion measure
\[
R(D)=\sum_{i=1}^m \frac{1}{2}\log\frac{\sigma_i^2}{D_i}.
\]
There are many practical vector quantization schemes \cite{Gersho&Gray:book}, including some of the machine learning techniques.

\paragraph{Decoder at Substations or Control Center}
The lower branch of Fig.~\ref{fig:IBC} describes the building blocks of the decoding process at the control center. For the most part, this process is the inverse of the upper branch, following mostly the procedure described in  \cite{Wang&Liu&Tong:21TPS}.
The vector quantization decoder reproduces the innovation sequence $\{(\hat{\nu}_i[n])\}$ for each subband (for those subbands that are not suppressed.)  The innovation autodecoder is used to produce the decompressed subband signal $\{(\hat{x}_i[n])\}$ (in the baseband form), and the subband decoder assembles the subband signal to form $\hat{x}[n]$ as the estimation of the original WMU data $x[n]$, for which the estimation error is controlled by $D$ in the rate-distortion function.

\paragraph{Central Data Analytics and Interactive Networking} Integrating data analytics with compression and networking provides the key component for interactive networking: the local IEDs can initiate the interaction based on local detection and transmit anomalous waveforms. Likewise, with the received data feeds from multiple IEDs, the central data analytics unit can initiate requests for high-resolution data when data feeds result in ambiguities and inference with low confidence. Such types of ``push-pull'' networking protocols have been widely used in streaming, Internet-of-Things, and content delivery networking technology \cite{Pathan&etal:08Bkchap}.

\section{Conclusion}
The future power grid will be dominated by power electronics-operated energy resources with fast dynamics and rich stochasticity. The existing grid monitoring and control framework is inadequate to address new challenges brought by large-scale integration of renewables, multi-sector electrification, and climate-change-induced natural disasters, and the next-generation monitoring and control need to include high-resolution synchro-waveform measurement technology and leverage powerful AI tools.

{The proposed physics-based AI foundation model is an initial attempt to extend the successful LLM-based foundation models to meet challenges in power system operations. Among the many challenges of this endeavor that have yet to be addressed here is the complexity involved in pre-training foundation models for power system applications. Unlike the LLM-based foundation model, where a single foundation model can be used for all text-based applications of the same language, a foundation model for power systems needs to be locational, as measurement at each location has its own characteristics and constraints. Although the innovation-based foundation model is considerably smaller than the LLM-based foundation model, the amount of pre-training can be significant. Using transfer learning techniques, as we have observed in our experiments, can significantly reduce the complexity of pre-training.}

\section*{Acknowledgement}
For this special issue in memory of the late Professor Bob Thomas, the first author (Lang Tong) wishes to acknowledge Professor Thomas's contribution to the work presented here over numerous conversations. Professor Thomas was pivotal in the first author's transition to teaching and research in renewable energy, electricity markets,  power systems, and smart grid. His presence at Cornell, the Power Systems Engineering Research Center (PSERC), and our research lab is sorely missed.

The authors also wish to acknowledge the contribution of Mei-Jen Lee who performed some of the simulations. We also thank the anonymous reviewers whose comments helped to improve the presentation of this work.

This work was supported in part by the National Science Foundation under Awards 2218110 and 2419622.

The authors have no relevant financial or non-financial interests to disclose.



\begin{thebibliography}{73}
\ifx \bisbn   \undefined \def \bisbn  #1{ISBN #1}\fi
\ifx \binits  \undefined \def \binits#1{#1}\fi
\ifx \bauthor  \undefined \def \bauthor#1{#1}\fi
\ifx \batitle  \undefined \def \batitle#1{#1}\fi
\ifx \bjtitle  \undefined \def \bjtitle#1{#1}\fi
\ifx \bvolume  \undefined \def \bvolume#1{\textbf{#1}}\fi
\ifx \byear  \undefined \def \byear#1{#1}\fi
\ifx \bissue  \undefined \def \bissue#1{#1}\fi
\ifx \bfpage  \undefined \def \bfpage#1{#1}\fi
\ifx \blpage  \undefined \def \blpage #1{#1}\fi
\ifx \burl  \undefined \def \burl#1{\textsf{#1}}\fi
\ifx \doiurl  \undefined \def \doiurl#1{\url{https://doi.org/#1}}\fi
\ifx \betal  \undefined \def \betal{\textit{et al.}}\fi
\ifx \binstitute  \undefined \def \binstitute#1{#1}\fi
\ifx \binstitutionaled  \undefined \def \binstitutionaled#1{#1}\fi
\ifx \bctitle  \undefined \def \bctitle#1{#1}\fi
\ifx \beditor  \undefined \def \beditor#1{#1}\fi
\ifx \bpublisher  \undefined \def \bpublisher#1{#1}\fi
\ifx \bbtitle  \undefined \def \bbtitle#1{#1}\fi
\ifx \bedition  \undefined \def \bedition#1{#1}\fi
\ifx \bseriesno  \undefined \def \bseriesno#1{#1}\fi
\ifx \blocation  \undefined \def \blocation#1{#1}\fi
\ifx \bsertitle  \undefined \def \bsertitle#1{#1}\fi
\ifx \bsnm \undefined \def \bsnm#1{#1}\fi
\ifx \bsuffix \undefined \def \bsuffix#1{#1}\fi
\ifx \bparticle \undefined \def \bparticle#1{#1}\fi
\ifx \barticle \undefined \def \barticle#1{#1}\fi
\bibcommenthead
\ifx \bconfdate \undefined \def \bconfdate #1{#1}\fi
\ifx \botherref \undefined \def \botherref #1{#1}\fi
\ifx \url \undefined \def \url#1{\textsf{#1}}\fi
\ifx \bchapter \undefined \def \bchapter#1{#1}\fi
\ifx \bbook \undefined \def \bbook#1{#1}\fi
\ifx \bcomment \undefined \def \bcomment#1{#1}\fi
\ifx \oauthor \undefined \def \oauthor#1{#1}\fi
\ifx \citeauthoryear \undefined \def \citeauthoryear#1{#1}\fi
\ifx \endbibitem  \undefined \def \endbibitem {}\fi
\ifx \bconflocation  \undefined \def \bconflocation#1{#1}\fi
\ifx \arxivurl  \undefined \def \arxivurl#1{\textsf{#1}}\fi
\csname PreBibitemsHook\endcsname

\bibitem[\protect\citeauthoryear{}{}]{ClimatMattes:2022}
\begin{botherref}
Surging Weather-related Power Outages.
\url{https://www.climatecentral.org/climate-matters/surging-weather-related-power-outages}.
Accessed:2024-03-08
\end{botherref}
\endbibitem

\bibitem[\protect\citeauthoryear{}{}]{Hussain:19}
\begin{botherref}
A Day Without Power: Outage Costs for Businesses.
\url{https://www.bloomenergy.com/blog/a-day-without-power-outage-costs-for-businesses/}.
Accessed: 2024-03-08
\end{botherref}
\endbibitem

\bibitem[\protect\citeauthoryear{Wiener}{1958}]{Wiener:58Book}
\begin{bbook}
\bauthor{\bsnm{Wiener}, \binits{N.}}:
\bbtitle{Nonlinear Problems in Random Theory}.
\bsertitle{M.I.T. paperback series}.
\bpublisher{MIT Press},
\blocation{Cambridge, MA}
(\byear{1958}).
\burl{https://books.google.com/books?id=HQBRAAAAMAAJ}
\end{bbook}
\endbibitem

\bibitem[\protect\citeauthoryear{Masani}{1966}]{Masani:66BAMS}
\begin{barticle}
\bauthor{\bsnm{Masani}, \binits{P.}}:
\batitle{{Wiener's contributions to generalized harmonic analysis, prediction
  theory and filter theory}}.
\bjtitle{Bulletin of the American Mathematical Society}
\bvolume{72}(\bissue{1.P2}),
\bfpage{73}--\blpage{125}
(\byear{1966})
\end{barticle}
\endbibitem

\bibitem[\protect\citeauthoryear{Mohsenian-Rad and
  Xu}{2023}]{Mohsenian-Rad&Xu:23}
\begin{barticle}
\bauthor{\bsnm{Mohsenian-Rad}, \binits{H.}},
\bauthor{\bsnm{Xu}, \binits{W.}}:
\batitle{Synchro-waveforms: A window to the future of power systems data
  analytics}.
\bjtitle{IEEE Power and Energy Magazine}
\bvolume{21}(\bissue{5}),
\bfpage{68}--\blpage{77}
(\byear{2023})
\end{barticle}
\endbibitem

\bibitem[\protect\citeauthoryear{Wang et~al.}{2021}]{Wang&Liu&Tong:21TPS}
\begin{barticle}
\bauthor{\bsnm{Wang}, \binits{X.}},
\bauthor{\bsnm{Liu}, \binits{Y.}},
\bauthor{\bsnm{Tong}, \binits{L.}}:
\batitle{Adaptive subband compression for streaming of continuous point-on-wave
  and pmu data}.
\bjtitle{IEEE Transactions on Power Systems}
\bvolume{36}(\bissue{6}),
\bfpage{5612}--\blpage{5621}
(\byear{2021})
\doiurl{10.1109/TPWRS.2021.3072882}
\end{barticle}
\endbibitem

\bibitem[\protect\citeauthoryear{Silverstein and
  Follum}{2020}]{Silverstein&Follum:20}
\begin{botherref}
\oauthor{\bsnm{Silverstein}, \binits{A.}},
\oauthor{\bsnm{Follum}, \binits{J.}}:
High-resolution, time-synchronized grid monitoring devices.
Technical Report NASPI-2020-TR-004,
North American Synchrophasor Initiative
(2020).
\url{https://www.naspi.org/sites/default/files/reference_documents/pnnl_29770_naspi_hires_synch_grid_devices_20200320.pdf}
\end{botherref}
\endbibitem

\bibitem[\protect\citeauthoryear{Anderson
  et~al.}{1999}]{Anderson&Agrawal&Ness:99}
\begin{bbook}
\bauthor{\bsnm{Anderson}, \binits{P.M.}},
\bauthor{\bsnm{Agrawal}, \binits{B.L.}},
\bauthor{\bsnm{{Van Ness}}, \binits{J.E.}}:
\bbtitle{Subsynchronous Resonance in Power Systems}.
\bpublisher{John Wiley \& Sons},
\blocation{New York}
(\byear{1999})
\end{bbook}
\endbibitem

\bibitem[\protect\citeauthoryear{Perez}{2010}]{Perez:10}
\begin{bchapter}
\bauthor{\bsnm{Perez}, \binits{J.}}:
\bctitle{A guide to digital fault recording event analysis}.
In: \bbtitle{Proc. 63rd Annu. Conf. Protective Relay Engineers},
pp. \bfpage{1}--\blpage{17}
(\byear{2010})
\end{bchapter}
\endbibitem

\bibitem[\protect\citeauthoryear{Cybenko and
  Brewington}{1999}]{Cybenko:99Bkchap}
\begin{bbook}
\bauthor{\bsnm{Cybenko}, \binits{G.}},
\bauthor{\bsnm{Brewington}, \binits{B.}}:
In: \beditor{\bsnm{Cybenko}, \binits{G.}},
\beditor{\bsnm{O'Leary}, \binits{D.P.}},
\beditor{\bsnm{Rissanen}, \binits{J.}} (eds.)
\bbtitle{The Foundations of Information Push and Pull},
pp. \bfpage{9}--\blpage{30}.
\bpublisher{Springer},
\blocation{New York, NY}
(\byear{1999}).
\doiurl{10.1007/978-1-4612-1524-0_2} .
\burl{https://doi.org/10.1007/978-1-4612-1524-0_2}
\end{bbook}
\endbibitem

\bibitem[\protect\citeauthoryear{Pathan et~al.}{2008}]{Pathan&etal:08Bkchap}
\begin{bbook}
\bauthor{\bsnm{Pathan}, \binits{M.}},
\bauthor{\bsnm{Buyya}, \binits{R.}},
\bauthor{\bsnm{Vakali}, \binits{A.}}:
In: \beditor{\bsnm{Buyya}, \binits{R.}},
\beditor{\bsnm{Pathan}, \binits{M.}},
\beditor{\bsnm{Vakali}, \binits{A.}} (eds.)
\bbtitle{Content Delivery Networks: State of the Art, Insights, and
  Imperatives},
pp. \bfpage{3}--\blpage{32}.
\bpublisher{Springer},
\blocation{Berlin, Heidelberg}
(\byear{2008}).
\doiurl{10.1007/978-3-540-77887-5_1} .
\burl{https://doi.org/10.1007/978-3-540-77887-5_1}
\end{bbook}
\endbibitem

\bibitem[\protect\citeauthoryear{Wischkaemper
  et~al.}{2015}]{Wischkaemper:15TSG}
\begin{barticle}
\bauthor{\bsnm{Wischkaemper}, \binits{J.A.}},
\bauthor{\bsnm{Benner}, \binits{C.L.}},
\bauthor{\bsnm{Russell}, \binits{B.D.}},
\bauthor{\bsnm{Manivannan}, \binits{K.}}:
\batitle{Application of waveform analytics for improved situational awareness
  of electric distribution feeders}.
\bjtitle{IEEE Transactions on Smart Grid}
\bvolume{6}(\bissue{4}),
\bfpage{2041}--\blpage{2049}
(\byear{2015})
\doiurl{10.1109/TSG.2015.2406757}
\end{barticle}
\endbibitem

\bibitem[\protect\citeauthoryear{Bastos et~al.}{2019}]{Bastos&etal:19PESGM}
\begin{bchapter}
\bauthor{\bsnm{Bastos}, \binits{A.F.}},
\bauthor{\bsnm{Santoso}, \binits{S.}},
\bauthor{\bsnm{Freitas}, \binits{W.}},
\bauthor{\bsnm{Xu}, \binits{W.}}:
\bctitle{Synchrowaveform measurement units and applications}.
In: \bbtitle{2019 IEEE Power \& Energy Society General Meetin G (PESGM)},
\bconflocation{Atlanta, GA, USA},
pp. \bfpage{1}--\blpage{5}
(\byear{2019}).
\doiurl{10.1109/PESGM40551.2019.8973736}
\end{bchapter}
\endbibitem

\bibitem[\protect\citeauthoryear{Carroll}{2019}]{Carroll19NAPSI}
\begin{botherref}
\oauthor{\bsnm{Carroll}, \binits{J.R.}}:
A Practical Approach to Streaming Point-on-Wave Data.
\url{https://www.naspi.org/sites/default/files/2019-04/04_gpa_carroll_practical_approach_pow_20190417.pdf}.
Accessed: {2024–03-09}
(2019)
\end{botherref}
\endbibitem

\bibitem[\protect\citeauthoryear{Rahmatian}{2019}]{Rahmatian:19}
\begin{botherref}
\oauthor{\bsnm{Rahmatian}, \binits{F.}}:
Point on Wave Measurements, Introduction.
\url{https://www.naspi.org/sites/default/files/2019-04/01_nugrid_rahmatian_pow_measurements_intro_20190416.pdf}.
Accessed: {2024–03-09}
(2019)
\end{botherref}
\endbibitem

\bibitem[\protect\citeauthoryear{Izadi and
  Mohsenian-Rad}{2020}]{Izadi&Mohsenian-Rad:20IGST}
\begin{bchapter}
\bauthor{\bsnm{Izadi}, \binits{M.}},
\bauthor{\bsnm{Mohsenian-Rad}, \binits{H.}}:
\bctitle{Event location identification in distribution networks using waveform
  measurement units}.
In: \bbtitle{2020 IEEE PES Innovative Smart Grid Technologies Europe
  (ISGT-Europe)},
\bconflocation{The Hague, Netherlands},
pp. \bfpage{924}--\blpage{928}
(\byear{2020}).
\doiurl{10.1109/NAPS58826.2023.10318763}
\end{bchapter}
\endbibitem

\bibitem[\protect\citeauthoryear{Xu et~al.}{2022}]{Xu&etal:22TPD}
\begin{barticle}
\bauthor{\bsnm{Xu}, \binits{W.}},
\bauthor{\bsnm{Huang}, \binits{Z.}},
\bauthor{\bsnm{Xie}, \binits{X.}},
\bauthor{\bsnm{Li}, \binits{C.}}:
\batitle{Synchronized waveforms – a frontier of data-based power system and
  apparatus monitoring, protection, and control}.
\bjtitle{IEEE Transactions on Power Delivery}
\bvolume{37}(\bissue{1}),
\bfpage{3}--\blpage{17}
(\byear{2022})
\doiurl{10.1109/TPWRD.2021.3072889}
\end{barticle}
\endbibitem

\bibitem[\protect\citeauthoryear{Rosenblatt}{1959}]{Rosenblatt:59}
\begin{barticle}
\bauthor{\bsnm{Rosenblatt}, \binits{M.}}:
\batitle{Stationary processes as shifts of functions of independent random
  variables}.
\bjtitle{Journal of Mathematics and Mechanics}
\bvolume{8}(\bissue{5}),
\bfpage{665}--\blpage{681}
(\byear{1959}).
Accessed 2024-01-18
\end{barticle}
\endbibitem

\bibitem[\protect\citeauthoryear{Bommasani et~al.}{2021}]{Bommasani&etal:22}
\begin{botherref}
\oauthor{\bsnm{Bommasani}, \binits{R.}},
\oauthor{\bsnm{Hudson}, \binits{D.A.}},
\oauthor{\bsnm{Adeli}, \binits{E.}},
\oauthor{\bsnm{Altman}, \binits{R.B.}},
\oauthor{\bsnm{Arora}, \binits{S.}},
\oauthor{\bsnm{Arx}, \binits{S.}},
\oauthor{\bsnm{Bernstein}, \binits{M.S.}},
\oauthor{\bsnm{Bohg}, \binits{J.}},
\oauthor{\bsnm{Bosselut}, \binits{A.}},
\oauthor{\bsnm{Brunskill}, \binits{E.}},
\oauthor{\bsnm{Brynjolfsson}, \binits{E.}},
\oauthor{\bsnm{Buch}, \binits{S.}},
\oauthor{\bsnm{Card}, \binits{D.}},
\oauthor{\bsnm{Castellon}, \binits{R.}},
\oauthor{\bsnm{Chatterji}, \binits{N.S.}},
\oauthor{\bsnm{Chen}, \binits{A.S.}},
\oauthor{\bsnm{Creel}, \binits{K.}},
\oauthor{\bsnm{Davis}, \binits{J.Q.}},
\oauthor{\bsnm{Demszky}, \binits{D.}},
\oauthor{\bsnm{Donahue}, \binits{C.}},
\oauthor{\bsnm{Doumbouya}, \binits{M.}},
\oauthor{\bsnm{Durmus}, \binits{E.}},
\oauthor{\bsnm{Ermon}, \binits{S.}},
\oauthor{\bsnm{Etchemendy}, \binits{J.}},
\oauthor{\bsnm{Ethayarajh}, \binits{K.}},
\oauthor{\bsnm{Fei{-}Fei}, \binits{L.}},
\oauthor{\bsnm{Finn}, \binits{C.}},
\oauthor{\bsnm{Gale}, \binits{T.}},
\oauthor{\bsnm{Gillespie}, \binits{L.E.}},
\oauthor{\bsnm{Goel}, \binits{K.}},
\oauthor{\bsnm{Goodman}, \binits{N.D.}},
\oauthor{\bsnm{Grossman}, \binits{S.}},
\oauthor{\bsnm{Guha}, \binits{N.}},
\oauthor{\bsnm{Hashimoto}, \binits{T.}},
\oauthor{\bsnm{Henderson}, \binits{P.}},
\oauthor{\bsnm{Hewitt}, \binits{J.}},
\oauthor{\bsnm{Ho}, \binits{D.E.}},
\oauthor{\bsnm{Hong}, \binits{J.}},
\oauthor{\bsnm{Hsu}, \binits{K.}},
\oauthor{\bsnm{Huang}, \binits{J.}},
\oauthor{\bsnm{Icard}, \binits{T.}},
\oauthor{\bsnm{Jain}, \binits{S.}},
\oauthor{\bsnm{Jurafsky}, \binits{D.}},
\oauthor{\bsnm{Kalluri}, \binits{P.}},
\oauthor{\bsnm{Karamcheti}, \binits{S.}},
\oauthor{\bsnm{Keeling}, \binits{G.}},
\oauthor{\bsnm{Khani}, \binits{F.}},
\oauthor{\bsnm{Khattab}, \binits{O.}},
\oauthor{\bsnm{Koh}, \binits{P.W.}},
\oauthor{\bsnm{Krass}, \binits{M.S.}},
\oauthor{\bsnm{Krishna}, \binits{R.}},
\oauthor{\bsnm{Kuditipudi}, \binits{R.}},
\oauthor{\bsnm{al.}}:
On the opportunities and risks of foundation models.
CoRR
\textbf{abs/2108.07258}
(2021)
{\href{https://arxiv.org/abs/2108.07258}{{2108.07258}}}
\end{botherref}
\endbibitem

\bibitem[\protect\citeauthoryear{Majumder et~al.}{2024}]{Majumder&etal:24Joule}
\begin{barticle}
\bauthor{\bsnm{Majumder}, \binits{S.}},
\bauthor{\bsnm{Dong}, \binits{L.}},
\bauthor{\bsnm{Doudi}, \binits{F.}},
\bauthor{\bsnm{Cai}, \binits{Y.}},
\bauthor{\bsnm{Tian}, \binits{C.}},
\bauthor{\bsnm{Kalathi}, \binits{D.}},
\bauthor{\bsnm{Ding}, \binits{K.}},
\bauthor{\bsnm{Thatte}, \binits{A.A.}},
\bauthor{\bsnm{Li}, \binits{N.}},
\bauthor{\bsnm{Xie}, \binits{L.}}:
\batitle{Exploring the capabilities and limitations of large language models in
  the electric energy sector}.
\bjtitle{Joule}
\bvolume{8}(\bissue{6}),
\bfpage{1544}--\blpage{1549}
(\byear{2024})
\end{barticle}
\endbibitem

\bibitem[\protect\citeauthoryear{Hamann et~al.}{2024}]{Hamann:24Joule}
\begin{barticle}
\bauthor{\bsnm{Hamann}, \binits{H.F.}},
\bauthor{\bsnm{Gjorgiev}, \binits{B.}},
\bauthor{\bsnm{Brunschwiler}, \binits{T.}},
\bauthor{\bsnm{Martins}, \binits{L.S.A.}},
\bauthor{\bsnm{Puech}, \binits{A.}},
\bauthor{\bsnm{Varbella}, \binits{A.}},
\bauthor{\bsnm{Weiss}, \binits{J.}},
\bauthor{\bsnm{Bernabe-Moreno}, \binits{J.}},
\bauthor{\bsnm{Massé}, \binits{A.B.}},
\bauthor{\bsnm{Choi}, \binits{S.L.}},
\bauthor{\bsnm{Foster}, \binits{I.}},
\bauthor{\bsnm{Hodge}, \binits{B.-M.}},
\bauthor{\bsnm{Jain}, \binits{R.}},
\bauthor{\bsnm{Kim}, \binits{K.}},
\bauthor{\bsnm{Mai}, \binits{V.}},
\bauthor{\bsnm{Mirallès}, \binits{F.}},
\bauthor{\bsnm{{De Montigny}}, \binits{M.}},
\bauthor{\bsnm{Ramos-Leaños}, \binits{O.}},
\bauthor{\bsnm{Suprême}, \binits{H.}},
\bauthor{\bsnm{Xie}, \binits{L.}},
\bauthor{\bsnm{Youssef}, \binits{E.-N.S.}},
\bauthor{\bsnm{Zinflou}, \binits{A.}},
\bauthor{\bsnm{Belyi}, \binits{A.}},
\bauthor{\bsnm{Bessa}, \binits{R.J.}},
\bauthor{\bsnm{Bhattarai}, \binits{B.P.}},
\bauthor{\bsnm{Schmude}, \binits{J.}},
\bauthor{\bsnm{Sobolevsky}, \binits{S.}}:
\batitle{Foundation models for the electric power grid}.
\bjtitle{Joule}
(\byear{2024})
\doiurl{10.1016/j.joule.2024.11.002}
\end{barticle}
\endbibitem

\bibitem[\protect\citeauthoryear{{U.S. Department of Energy}}{2024}]{DOE24_rpt}
\begin{botherref}
\oauthor{\bsnm{{U.S. Department of Energy}}}:
AI for Energy: Opportunities for a Modern Grid and Clean Energy Economy.
\url{https://www.energy.gov/sites/default/files/2024-04/AI\%20EO\%20Report\%20Section\%205.2g\%28i\%29_043024.pdf}
\end{botherref}
\endbibitem

\bibitem[\protect\citeauthoryear{Wang et~al.}{2024}]{Wang&Tong&Zhao:24arxiv}
\begin{botherref}
\oauthor{\bsnm{Wang}, \binits{X.}},
\oauthor{\bsnm{Tong}, \binits{L.}},
\oauthor{\bsnm{Zhao}, \binits{Q.}}:
{Generative Probabilistic Price Forecasting via Weak Innovations}.
Submitted for publications. See updated preprint at {\tt arXiv}
(2024)
\end{botherref}
\endbibitem

\bibitem[\protect\citeauthoryear{Rosenblatt}{2009}]{Rosenblatt:09}
\begin{barticle}
\bauthor{\bsnm{Rosenblatt}, \binits{M.}}:
\batitle{A comment on a conjecture of n. wiener}.
\bjtitle{Statistics \& Probability Letters}
\bvolume{79}(\bissue{3}),
\bfpage{347}--\blpage{348}
(\byear{2009})
\doiurl{10.1016/j.spl.2008.09.001}
\end{barticle}
\endbibitem

\bibitem[\protect\citeauthoryear{Wu}{2005}]{Wu:05PNAS}
\begin{barticle}
\bauthor{\bsnm{Wu}, \binits{W.}}:
\batitle{{Nonlinear System Theory: Another Look at Dependence}}.
\bjtitle{Proceedings of National Academy of Sciences}
\bvolume{102}(\bissue{40}),
\bfpage{14150}--\blpage{14154}
(\byear{2005})
\end{barticle}
\endbibitem

\bibitem[\protect\citeauthoryear{Wu}{2011}]{Wu:11}
\begin{barticle}
\bauthor{\bsnm{Wu}, \binits{W.}}:
\batitle{{Asymptotic Theory for Stationary Processes}}.
\bjtitle{Statistics and Its Interface}
\bvolume{0},
\bfpage{1}--\blpage{20}
(\byear{2011})
\doiurl{10.4310/SII.2011.v4.n2.a15}
\end{barticle}
\endbibitem

\bibitem[\protect\citeauthoryear{Wang and Tong}{2021}]{Wang&Tong:21JMLR}
\begin{botherref}
\oauthor{\bsnm{Wang}, \binits{X.}},
\oauthor{\bsnm{Tong}, \binits{L.}}:
{Innovations Autoencoder and its Application in One-class Anomalous Sequence
  Detection}.
arXiv:2106.12382
(2021).
\url{https://arxiv.org/abs/2106.12382}
\end{botherref}
\endbibitem

\bibitem[\protect\citeauthoryear{Paszke
  et~al.}{2019}]{paszke2019pytorchimperativestylehighperformance}
\begin{botherref}
\oauthor{\bsnm{Paszke}, \binits{A.}},
\oauthor{\bsnm{Gross}, \binits{S.}},
\oauthor{\bsnm{Massa}, \binits{F.}},
\oauthor{\bsnm{Lerer}, \binits{A.}},
\oauthor{\bsnm{Bradbury}, \binits{J.}},
\oauthor{\bsnm{Chanan}, \binits{G.}},
\oauthor{\bsnm{Killeen}, \binits{T.}},
\oauthor{\bsnm{Lin}, \binits{Z.}},
\oauthor{\bsnm{Gimelshein}, \binits{N.}},
\oauthor{\bsnm{Antiga}, \binits{L.}},
\oauthor{\bsnm{Desmaison}, \binits{A.}},
\oauthor{\bsnm{Köpf}, \binits{A.}},
\oauthor{\bsnm{Yang}, \binits{E.}},
\oauthor{\bsnm{DeVito}, \binits{Z.}},
\oauthor{\bsnm{Raison}, \binits{M.}},
\oauthor{\bsnm{Tejani}, \binits{A.}},
\oauthor{\bsnm{Chilamkurthy}, \binits{S.}},
\oauthor{\bsnm{Steiner}, \binits{B.}},
\oauthor{\bsnm{Fang}, \binits{L.}},
\oauthor{\bsnm{Bai}, \binits{J.}},
\oauthor{\bsnm{Chintala}, \binits{S.}}:
PyTorch: An Imperative Style, High-Performance Deep Learning Library
(2019).
\url{https://arxiv.org/abs/1912.01703}
\end{botherref}
\endbibitem

\bibitem[\protect\citeauthoryear{Abadi
  et~al.}{2015}]{tensorflow2015-whitepaper}
\begin{botherref}
\oauthor{\bsnm{Abadi}, \binits{M.}},
\oauthor{\bsnm{Agarwal}, \binits{A.}},
\oauthor{\bsnm{Barham}, \binits{P.}},
\oauthor{\bsnm{Brevdo}, \binits{E.}},
\oauthor{\bsnm{Chen}, \binits{Z.}},
\oauthor{\bsnm{Citro}, \binits{C.}},
\oauthor{\bsnm{Corrado}, \binits{G.S.}},
\oauthor{\bsnm{Davis}, \binits{A.}},
\oauthor{\bsnm{Dean}, \binits{J.}},
\oauthor{\bsnm{Devin}, \binits{M.}},
\oauthor{\bsnm{Ghemawat}, \binits{S.}},
\oauthor{\bsnm{Goodfellow}, \binits{I.}},
\oauthor{\bsnm{Harp}, \binits{A.}},
\oauthor{\bsnm{Irving}, \binits{G.}},
\oauthor{\bsnm{Isard}, \binits{M.}},
\oauthor{\bsnm{Jia}, \binits{Y.}},
\oauthor{\bsnm{Jozefowicz}, \binits{R.}},
\oauthor{\bsnm{Kaiser}, \binits{L.}},
\oauthor{\bsnm{Kudlur}, \binits{M.}},
\oauthor{\bsnm{Levenberg}, \binits{J.}},
\oauthor{\bsnm{Man\'{e}}, \binits{D.}},
\oauthor{\bsnm{Monga}, \binits{R.}},
\oauthor{\bsnm{Moore}, \binits{S.}},
\oauthor{\bsnm{Murray}, \binits{D.}},
\oauthor{\bsnm{Olah}, \binits{C.}},
\oauthor{\bsnm{Schuster}, \binits{M.}},
\oauthor{\bsnm{Shlens}, \binits{J.}},
\oauthor{\bsnm{Steiner}, \binits{B.}},
\oauthor{\bsnm{Sutskever}, \binits{I.}},
\oauthor{\bsnm{Talwar}, \binits{K.}},
\oauthor{\bsnm{Tucker}, \binits{P.}},
\oauthor{\bsnm{Vanhoucke}, \binits{V.}},
\oauthor{\bsnm{Vasudevan}, \binits{V.}},
\oauthor{\bsnm{Vi\'{e}gas}, \binits{F.}},
\oauthor{\bsnm{Vinyals}, \binits{O.}},
\oauthor{\bsnm{Warden}, \binits{P.}},
\oauthor{\bsnm{Wattenberg}, \binits{M.}},
\oauthor{\bsnm{Wicke}, \binits{M.}},
\oauthor{\bsnm{Yu}, \binits{Y.}},
\oauthor{\bsnm{Zheng}, \binits{X.}}:
{TensorFlow}: Large-Scale Machine Learning on Heterogeneous Systems.
Software available from tensorflow.org
(2015).
\url{https://www.tensorflow.org/}
\end{botherref}
\endbibitem

\bibitem[\protect\citeauthoryear{Sheskin}{2011}]{Sheskin:11}
\begin{bbook}
\bauthor{\bsnm{Sheskin}, \binits{D.J.}}:
\bbtitle{Handbook of Parametric and Nonparametric Statistical Procedures, Fifth
  Editio (5th Ed.)}.
\bpublisher{Chapman and Hall/CRC},
\blocation{Boca Raton}
(\byear{2011}).
\burl{https://doi.org/10.1201/9780429186196}
\end{bbook}
\endbibitem

\bibitem[\protect\citeauthoryear{Sheskin}{2007}]{Sheskin:07book}
\begin{bbook}
\bauthor{\bsnm{Sheskin}, \binits{D.J.}}:
\bbtitle{Handbook of Parametric and Nonparametric Statistical Procedures},
\bedition{4th} edn.,
p. \bfpage{1193}.
\bpublisher{Chapman \& Hall/CRC}, \blocation{Boca Raton}
(\byear{2007})
\end{bbook}
\endbibitem

\bibitem[\protect\citeauthoryear{Ma and Perkins}{2003}]{Ma&Perkins:03IJCNN}
\begin{bchapter}
\bauthor{\bsnm{Ma}, \binits{J.}},
\bauthor{\bsnm{Perkins}, \binits{S.}}:
\bctitle{{Time-series Novelty Detection Using One-class Support Vector
  Machines}}.
In: \bbtitle{Proceedings of the International Joint Conference on Neural
  Networks, 2003.},
vol. \bseriesno{3},
pp. \bfpage{1741}--\blpage{17453}
(\byear{2003}).
\doiurl{10.1109/IJCNN.2003.1223670}
\end{bchapter}
\endbibitem

\bibitem[\protect\citeauthoryear{Dasgupta and
  Forrest}{1995}]{Dasgupta&Forrest:1995ICIS}
\begin{bchapter}
\bauthor{\bsnm{Dasgupta}, \binits{D.}},
\bauthor{\bsnm{Forrest}, \binits{S.}}:
\bctitle{Novelty {Detection} in {Time} {Series} {Data} using {Ideas} from
  {Immunology}}.
In: \bbtitle{In {Proceedings} of {The} {International} {Conference} on
  {Intelligent} {Systems}}
(\byear{1995})
\end{bchapter}
\endbibitem

\bibitem[\protect\citeauthoryear{Gardner et~al.}{2006}]{Gardner&etal:2006JMLR}
\begin{barticle}
\bauthor{\bsnm{Gardner}, \binits{A.B.}},
\bauthor{\bsnm{Krieger}, \binits{A.M.}},
\bauthor{\bsnm{Vachtsevanos}, \binits{G.}},
\bauthor{\bsnm{Litt}, \binits{B.}}:
\batitle{One-{Class} {Novelty} {Detection} for {Seizure} {Analysis} from
  {Intracranial} {EEG}}.
\bjtitle{Journal of Machine Learning Research}
\bvolume{7}(\bissue{37}),
\bfpage{1025}--\blpage{1044}
(\byear{2006})
\end{barticle}
\endbibitem

\bibitem[\protect\citeauthoryear{Sch\"{o}lkopf et~al.}{1999}]{Scholkopf:99NIPS}
\begin{botherref}
\oauthor{\bsnm{Sch\"{o}lkopf}, \binits{B.}},
\oauthor{\bsnm{Williamson}, \binits{R.}},
\oauthor{\bsnm{Smola}, \binits{A.}},
\oauthor{\bsnm{Shawe-Taylor}, \binits{J.}},
\oauthor{\bsnm{Platt}, \binits{J.}}:
{Support Vector Method for Novelty Detection}.
Proceedings of the 12th International Conference on Neural Information
  Processing Systems,
582--588
(1999)
\end{botherref}
\endbibitem

\bibitem[\protect\citeauthoryear{Khan and Madden}{2014}]{Khan&Madden:04}
\begin{barticle}
\bauthor{\bsnm{Khan}, \binits{S.S.}},
\bauthor{\bsnm{Madden}, \binits{M.G.}}:
\batitle{{One-class Classification: Taxonomy of Study and Review of
  Techniques}}.
\bjtitle{The Knowledge Engineering Review}
\bvolume{29}(\bissue{3}),
\bfpage{345}--\blpage{374}
(\byear{2014})
\doiurl{10.1017/S026988891300043X}
\end{barticle}
\endbibitem

\bibitem[\protect\citeauthoryear{Bergmann et~al.}{2019}]{Bergmann&etal:19}
\begin{barticle}
\bauthor{\bsnm{Bergmann}, \binits{P.}},
\bauthor{\bsnm{Löwe}, \binits{S.}},
\bauthor{\bsnm{Fauser}, \binits{M.}},
\bauthor{\bsnm{Sattlegger}, \binits{D.}},
\bauthor{\bsnm{Steger}, \binits{C.}}:
\batitle{{Improving Unsupervised Defect Segmentation by Applying Structural
  Similarity to Autoencoders}}.
\bjtitle{Proceedings of the 14th International Joint Conference on Computer
  Vision, Imaging and Computer Graphics Theory and Applications}
(\byear{2019})
\doiurl{10.5220/0007364503720380}
\end{barticle}
\endbibitem

\bibitem[\protect\citeauthoryear{Gong et~al.}{2019}]{Gong&etal:19}
\begin{botherref}
\oauthor{\bsnm{Gong}, \binits{D.}},
\oauthor{\bsnm{Liu}, \binits{L.}},
\oauthor{\bsnm{Le}, \binits{V.}},
\oauthor{\bsnm{Saha}, \binits{B.}},
\oauthor{\bsnm{Mansour}, \binits{M.R.}},
\oauthor{\bsnm{Venkatesh}, \binits{S.}},
\oauthor{\bsnm{Hengel}, \binits{A.}}:
{Memorizing Normality to Detect Anomaly: Memory-augmented Deep Autoencoder for
  Unsupervised Anomaly Detection}
(2019)
\end{botherref}
\endbibitem

\bibitem[\protect\citeauthoryear{Lee et~al.}{2018}]{Lee&etal:18ICLR}
\begin{bchapter}
\bauthor{\bsnm{Lee}, \binits{K.}},
\bauthor{\bsnm{Lee}, \binits{H.}},
\bauthor{\bsnm{Lee}, \binits{K.}},
\bauthor{\bsnm{Shin}, \binits{J.}}:
\bctitle{{Training Confidence-calibrated Classifiers for Detecting
  Out-of-Distribution Samples}}.
In: \bbtitle{International Conference on Learning Representations}
(\byear{2018}).
\burl{https://openreview.net/forum?id=ryiAv2xAZ}
\end{bchapter}
\endbibitem

\bibitem[\protect\citeauthoryear{Hendrycks
  et~al.}{2019}]{Hendrycks&etal:18ICLR}
\begin{bchapter}
\bauthor{\bsnm{Hendrycks}, \binits{D.}},
\bauthor{\bsnm{Mazeika}, \binits{M.}},
\bauthor{\bsnm{Dietterich}, \binits{T.}}:
\bctitle{{Deep Anomaly Detection with Outlier Exposure}}.
In: \bbtitle{International Conference on Learning Representations}
(\byear{2019}).
\burl{https://openreview.net/forum?id=HyxCxhRcY7}
\end{bchapter}
\endbibitem

\bibitem[\protect\citeauthoryear{Ren et~al.}{2019}]{Ren&etal:19ICLR}
\begin{botherref}
\oauthor{\bsnm{Ren}, \binits{J.}},
\oauthor{\bsnm{Liu}, \binits{P.J.}},
\oauthor{\bsnm{Fertig}, \binits{E.}},
\oauthor{\bsnm{Snoek}, \binits{J.}},
\oauthor{\bsnm{Poplin}, \binits{R.}},
\oauthor{\bsnm{DePristo}, \binits{M.A.}},
\oauthor{\bsnm{Dillon}, \binits{J.V.}},
\oauthor{\bsnm{Lakshminarayanan}, \binits{B.}}:
{Likelihood Ratios for Out-of-Distribution Detection}.
arXiv:1906.02845
(2019).
\url{https://arxiv.org/abs/1906.02845}
\end{botherref}
\endbibitem

\bibitem[\protect\citeauthoryear{Hendrycks and
  Gimpel}{2017}]{Hendrycks&Gimpel:17ICLR}
\begin{bchapter}
\bauthor{\bsnm{Hendrycks}, \binits{D.}},
\bauthor{\bsnm{Gimpel}, \binits{K.}}:
\bctitle{A baseline for detecting misclassified and out-of-distribution
  examples in neural networks}.
In: \bbtitle{International Conference on Learning Representations}
(\byear{2017}).
\burl{https://openreview.net/forum?id=Hkg4TI9xl}
\end{bchapter}
\endbibitem

\bibitem[\protect\citeauthoryear{Lakshminarayanan
  et~al.}{2017}]{Lakshminarayanan&Pritzel&Blundell:17NIPS}
\begin{bchapter}
\bauthor{\bsnm{Lakshminarayanan}, \binits{B.}},
\bauthor{\bsnm{Pritzel}, \binits{A.}},
\bauthor{\bsnm{Blundell}, \binits{C.}}:
\bctitle{Simple and scalable predictive uncertainty estimation using deep
  ensembles}.
In: \beditor{\bsnm{Guyon}, \binits{I.}},
\beditor{\bsnm{Luxburg}, \binits{U.V.}},
\beditor{\bsnm{Bengio}, \binits{S.}},
\beditor{\bsnm{Wallach}, \binits{H.}},
\beditor{\bsnm{Fergus}, \binits{R.}},
\beditor{\bsnm{Vishwanathan}, \binits{S.}},
\beditor{\bsnm{Garnett}, \binits{R.}} (eds.)
\bbtitle{Advances in Neural Information Processing Systems},
vol. \bseriesno{30}.
\bpublisher{Curran Associates, Inc.},
\blocation{New Orleans, Louisiana, USA}
(\byear{2017}).
\burl{https://proceedings.neurips.cc/paper_files/paper/2017/file/9ef2ed4b7fd2c810847ffa5fa85bce38-Paper.pdf}
\end{bchapter}
\endbibitem

\bibitem[\protect\citeauthoryear{Liang et~al.}{2018}]{Liang&Li&Srikant:18ICLR}
\begin{bchapter}
\bauthor{\bsnm{Liang}, \binits{S.}},
\bauthor{\bsnm{Li}, \binits{Y.}},
\bauthor{\bsnm{Srikant}, \binits{R.}}:
\bctitle{Enhancing the Reliability of Out-of-distribution Image Detection in
  Neural Networks}.
(\byear{2018}).
\bcomment{Funding Information: The research reported here was supported by NSF
  Grant CPS ECCS 1739189. Publisher Copyright: {\textcopyright} Learning
  Representations, ICLR 2018 - Conference Track Proceedings.All right
  reserved.; 6th International Conference on Learning Representations, ICLR
  2018 ; Conference date: 30-04-2018 Through 03-05-2018}
\end{bchapter}
\endbibitem

\bibitem[\protect\citeauthoryear{Lee et~al.}{2018}]{Lee&etal:18NIPS}
\begin{bchapter}
\bauthor{\bsnm{Lee}, \binits{K.}},
\bauthor{\bsnm{Lee}, \binits{K.}},
\bauthor{\bsnm{Lee}, \binits{H.}},
\bauthor{\bsnm{Shin}, \binits{J.}}:
\bctitle{{A Simple Unified Framework for Detecting Out-of-Distribution Samples
  and Adversarial Attacks}}.
In: \beditor{\bsnm{Bengio}, \binits{S.}},
\beditor{\bsnm{Wallach}, \binits{H.}},
\beditor{\bsnm{Larochelle}, \binits{H.}},
\beditor{\bsnm{Grauman}, \binits{K.}},
\beditor{\bsnm{Cesa-Bianchi}, \binits{N.}},
\beditor{\bsnm{Garnett}, \binits{R.}} (eds.)
\bbtitle{Advances in Neural Information Processing Systems},
vol. \bseriesno{31}.
\bpublisher{Curran Associates, Inc.},
\blocation{New Orleans, Louisiana, USA}
(\byear{2018}).
\burl{https://proceedings.neurips.cc/paper/2018/file/abdeb6f575ac5c6676b747bca8d09cc2-Paper.pdf}
\end{bchapter}
\endbibitem

\bibitem[\protect\citeauthoryear{Le~Lan and Dinh}{2021}]{Lan&Dinh:21}
\begin{botherref}
\oauthor{\bsnm{Le~Lan}, \binits{C.}},
\oauthor{\bsnm{Dinh}, \binits{L.}}:
Perfect density models cannot guarantee anomaly detection.
Entropy
\textbf{23}(12)
(2021)
\doiurl{10.3390/e23121690}
\end{botherref}
\endbibitem

\bibitem[\protect\citeauthoryear{Schlegl et~al.}{2019}]{Schlegl&Seebock:19}
\begin{barticle}
\bauthor{\bsnm{Schlegl}, \binits{T.}},
\bauthor{\bsnm{Seeböck}, \binits{P.}},
\bauthor{\bsnm{Waldstein}, \binits{S.M.}},
\bauthor{\bsnm{Langs}, \binits{G.}},
\bauthor{\bsnm{Schmidt-Erfurth}, \binits{U.}}:
\batitle{f-anogan: Fast unsupervised anomaly detection with generative
  adversarial networks}.
\bjtitle{Medical Image Analysis}
\bvolume{54},
\bfpage{30}--\blpage{44}
(\byear{2019})
\doiurl{10.1016/j.media.2019.01.010}
\end{barticle}
\endbibitem

\bibitem[\protect\citeauthoryear{Dinh et~al.}{2015}]{Dinh&Krueger&Bengio:2015}
\begin{botherref}
\oauthor{\bsnm{Dinh}, \binits{L.}},
\oauthor{\bsnm{Krueger}, \binits{D.}},
\oauthor{\bsnm{Bengio}, \binits{Y.}}:
{{NICE:} Non-linear Independent Components Estimation}.
arXiv:arXiv:1410.8516
(2015).
\url{https://arxiv.org/abs/1410.8516}
\end{botherref}
\endbibitem

\bibitem[\protect\citeauthoryear{Brakel and Bengio}{2017}]{Brakel&Bengio:17}
\begin{botherref}
\oauthor{\bsnm{Brakel}, \binits{P.}},
\oauthor{\bsnm{Bengio}, \binits{Y.}}:
{Learning Independent Features with Adversarial Nets for Non-linear ICA}.
arXiv:1710.05050
(2017)
\end{botherref}
\endbibitem

\bibitem[\protect\citeauthoryear{Zhou et~al.}{2021}]{Zhou&etal:21AAAI}
\begin{barticle}
\bauthor{\bsnm{Zhou}, \binits{H.}},
\bauthor{\bsnm{Zhang}, \binits{S.}},
\bauthor{\bsnm{Peng}, \binits{J.}},
\bauthor{\bsnm{Zhang}, \binits{S.}},
\bauthor{\bsnm{Li}, \binits{J.}},
\bauthor{\bsnm{Xiong}, \binits{H.}},
\bauthor{\bsnm{Zhang}, \binits{W.}}:
\batitle{Informer: Beyond efficient transformer for long sequence time-series
  forecasting}.
\bjtitle{Proceedings of the AAAI Conference on Artificial Intelligence}
\bvolume{35}(\bissue{12}),
\bfpage{11106}--\blpage{11115}
(\byear{2021})
\doiurl{10.1609/aaai.v35i12.17325}
\end{barticle}
\endbibitem

\bibitem[\protect\citeauthoryear{Zeng et~al.}{2023}]{Zeng&etal:23IAAI}
\begin{bchapter}
\bauthor{\bsnm{Zeng}, \binits{A.}},
\bauthor{\bsnm{Chen}, \binits{M.}},
\bauthor{\bsnm{Zhang}, \binits{L.}},
\bauthor{\bsnm{Xu}, \binits{Q.}}:
\bctitle{Are transformers effective for time series forecasting?}
In: \bbtitle{Proceedings of the Thirty-Seventh AAAI Conference on Artificial
  Intelligence and Thirty-Fifth Conference on Innovative Applications of
  Artificial Intelligence and Thirteenth Symposium on Educational Advances in
  Artificial Intelligence}.
\bsertitle{AAAI'23/IAAI'23/EAAI'23}.
\bpublisher{AAAI Press},
(\byear{2023}).
\doiurl{10.1609/aaai.v37i9.26317} .
\burl{https://doi.org/10.1609/aaai.v37i9.26317}
\end{bchapter}
\endbibitem

\bibitem[\protect\citeauthoryear{D'Agostino}{2017}]{GOFbook}
\begin{bbook}
\bauthor{\bsnm{D'Agostino}, \binits{R.B.}}:
\bbtitle{Goodness-of-Fit-Techniques. (First Edition)}.
\bpublisher{Taylor and Francis},
\blocation{London}
(\byear{2017}).
\burl{https://www.taylorfrancis.com/books/9780203753064}
\end{bbook}
\endbibitem

\bibitem[\protect\citeauthoryear{Neyman}{1937}]{Neyman:37SAJ}
\begin{barticle}
\bauthor{\bsnm{Neyman}, \binits{J.}}:
\batitle{"smooth test" for goodness of fit}.
\bjtitle{Scandinavian Actuarial Journal}
\bvolume{1937}(\bissue{3-4}),
\bfpage{149}--\blpage{199}
(\byear{1937})
\doiurl{10.1080/03461238.1937.10404821}
{\href{https://arxiv.org/abs/https://doi.org/10.1080/03461238.1937.10404821}{{https://doi.org/10.1080/03461238.1937.10404821}}}
\end{barticle}
\endbibitem

\bibitem[\protect\citeauthoryear{Rayner and Best}{1990}]{Rayner&Best:90}
\begin{barticle}
\bauthor{\bsnm{Rayner}, \binits{J.C.W.}},
\bauthor{\bsnm{Best}, \binits{D.J.}}:
\batitle{Smooth tests of goodness of fit: An overview}.
\bjtitle{International Statistical Review / Revue Internationale de
  Statistique}
\bvolume{58}(\bissue{1}),
\bfpage{9}--\blpage{17}
(\byear{1990}).
Accessed 2024-01-18
\end{barticle}
\endbibitem

\bibitem[\protect\citeauthoryear{Fawzi et~al.}{2022}]{OufkirEtalNeurips2021}
\begin{botherref}
\oauthor{\bsnm{Fawzi}, \binits{O.}},
\oauthor{\bsnm{Flammarion}, \binits{N.}},
\oauthor{\bsnm{Garivier}, \binits{A.}},
\oauthor{\bsnm{Oufkir}, \binits{A.}}:
Sequential algorithms for testing identity and closeness of distributions
(2022)
\end{botherref}
\endbibitem

\bibitem[\protect\citeauthoryear{Telukunta
  et~al.}{2017}]{TelukuntaEtal17:CSEEPES}
\begin{barticle}
\bauthor{\bsnm{Telukunta}, \binits{V.}},
\bauthor{\bsnm{Pradhan}, \binits{J.}},
\bauthor{\bsnm{Agrawal}, \binits{A.}},
\bauthor{\bsnm{Singh}, \binits{M.}},
\bauthor{\bsnm{Srivani}, \binits{S.G.}}:
\batitle{{Protection Challenges under Bulk Penetration of Renewable Energy
  Resources in Power Systems: A Review}}.
\bjtitle{CSEE Journal of Power and Energy Systems}
\bvolume{3}(\bissue{4}),
\bfpage{365}--\blpage{379}
(\byear{2017})
\doiurl{10.17775/CSEEJPES.2017.00030}
\end{barticle}
\endbibitem

\bibitem[\protect\citeauthoryear{Anderson}{1999}]{Anderson99}
\begin{bbook}
\bauthor{\bsnm{Anderson}, \binits{P.M.}}:
\bbtitle{Power System Protection}.
\bpublisher{Mcgraw-Hill},
\blocation{New York}
(\byear{1999})
\end{bbook}
\endbibitem

\bibitem[\protect\citeauthoryear{El-Khattam and
  Sidhu}{2008}]{Elkhatam&Sidhu08TPD}
\begin{barticle}
\bauthor{\bsnm{El-Khattam}, \binits{W.}},
\bauthor{\bsnm{Sidhu}, \binits{T.S.}}:
\batitle{{Restoration of Directional Overcurrent Relay Coordination in
  Distributed Generation Systems Utilizing Fault Current Limiter}}.
\bjtitle{IEEE Transactions on Power Delivery}
\bvolume{23}(\bissue{2}),
\bfpage{576}--\blpage{585}
(\byear{2008})
\doiurl{10.1109/TPWRD.2008.915778}
\end{barticle}
\endbibitem

\bibitem[\protect\citeauthoryear{Ibrahim et~al.}{2017}]{Ibrahim17IJEPES}
\begin{barticle}
\bauthor{\bsnm{Ibrahim}, \binits{D.K.}},
\bauthor{\bsnm{Zahab}, \binits{E.}},
\bauthor{\bsnm{Mostafa}, \binits{S.}}:
\batitle{{New Coordination Approach to Minimize the Number of Re-adjusted
  Relays when Adding DGs in Interconnected Power Systems with a Minimum Value
  of Fault Current Limiter}}.
\bjtitle{International Journal of Electrical Power \& Energy Systems}
\bvolume{85},
\bfpage{32}--\blpage{41}
(\byear{2017})
\doiurl{10.1016/j.ijepes.2016.08.003}
\end{barticle}
\endbibitem

\bibitem[\protect\citeauthoryear{Chattopadhyay
  et~al.}{1996}]{ChattopadhyaySachdevSidhu96:TPD}
\begin{barticle}
\bauthor{\bsnm{Chattopadhyay}, \binits{B.}},
\bauthor{\bsnm{Sachdev}, \binits{M.S.}},
\bauthor{\bsnm{Sidhu}, \binits{T.S.}}:
\batitle{{An On-line Relay Coordination Algorithm for Adaptive Protection using
  Linear Programming Technique}}.
\bjtitle{IEEE Transactions on Power Delivery}
\bvolume{11}(\bissue{1}),
\bfpage{165}--\blpage{173}
(\byear{1996})
\doiurl{10.1109/61.484013}
\end{barticle}
\endbibitem

\bibitem[\protect\citeauthoryear{Liu et~al.}{2016}]{LiuEtal:16CSEEPES}
\begin{barticle}
\bauthor{\bsnm{Liu}, \binits{Z.}},
\bauthor{\bsnm{Hoidalen}, \binits{H.K.}},
\bauthor{\bsnm{Saha}, \binits{M.M.}}:
\batitle{{An Intelligent Coordinated Protection and Control Strategy for
  Distribution Network with Wind Generation Integration}}.
\bjtitle{CSEE Journal of Power and Energy Systems}
\bvolume{2}(\bissue{4}),
\bfpage{23}--\blpage{30}
(\byear{2016})
\doiurl{10.17775/CSEEJPES.2016.00045}
\end{barticle}
\endbibitem

\bibitem[\protect\citeauthoryear{Wan et~al.}{2010}]{WanLiWong10:TIA}
\begin{barticle}
\bauthor{\bsnm{Wan}, \binits{H.}},
\bauthor{\bsnm{Li}, \binits{K.K.}},
\bauthor{\bsnm{Wong}, \binits{K.P.}}:
\batitle{{An Adaptive Multiagent Approach to Protection Relay Coordination With
  Distributed Generators in Industrial Power Distribution System}}.
\bjtitle{IEEE Transactions on Industry Applications}
\bvolume{46}(\bissue{5}),
\bfpage{2118}--\blpage{2124}
(\byear{2010})
\doiurl{10.1109/TIA.2010.2059492}
\end{barticle}
\endbibitem

\bibitem[\protect\citeauthoryear{Papaspiliotopoulos
  et~al.}{2017}]{PapaspiliotopoulosEtal:17TPD}
\begin{barticle}
\bauthor{\bsnm{Papaspiliotopoulos}, \binits{V.A.}},
\bauthor{\bsnm{Korres}, \binits{G.N.}},
\bauthor{\bsnm{Kleftakis}, \binits{V.A.}},
\bauthor{\bsnm{Hatziargyriou}, \binits{N.D.}}:
\batitle{{Hardware-In-the-Loop Design and Optimal Setting of Adaptive
  Protection Schemes for Distribution Systems With Distributed Generation}}.
\bjtitle{IEEE Transactions on Power Delivery}
\bvolume{32}(\bissue{1}),
\bfpage{393}--\blpage{400}
(\byear{2017})
\doiurl{10.1109/TPWRD.2015.2509784}
\end{barticle}
\endbibitem

\bibitem[\protect\citeauthoryear{Rezaei}{2019}]{Rezaei19EEEIC}
\begin{bchapter}
\bauthor{\bsnm{Rezaei}, \binits{S.}}:
\bctitle{{Intelligent Overcurrent Protection During Ferroresonance in Smart
  Distribution Grid}}.
In: \bbtitle{2019 IEEE International Conference on Environment and Electrical
  Engineering and 2019 IEEE Industrial and Commercial Power Systems Europe
  (EEEIC / I CPS Europe)},
pp. \bfpage{1}--\blpage{6}
(\byear{2019}).
\doiurl{10.1109/EEEIC.2019.8783752}
\end{bchapter}
\endbibitem

\bibitem[\protect\citeauthoryear{Shen et~al.}{2017}]{ShenEtal:17TPD}
\begin{barticle}
\bauthor{\bsnm{Shen}, \binits{S.}},
\bauthor{\bsnm{Lin}, \binits{D.}},
\bauthor{\bsnm{Wang}, \binits{H.}},
\bauthor{\bsnm{Hu}, \binits{P.}},
\bauthor{\bsnm{Jiang}, \binits{K.}},
\bauthor{\bsnm{Lin}, \binits{D.}},
\bauthor{\bsnm{He}, \binits{B.}}:
\batitle{{An Adaptive Protection Scheme for Distribution Systems With DGs Based
  on Optimized Thevenin Equivalent Parameters Estimation}}.
\bjtitle{IEEE Transactions on Power Delivery}
\bvolume{32}(\bissue{1}),
\bfpage{411}--\blpage{419}
(\byear{2017})
\doiurl{10.1109/TPWRD.2015.2506155}
\end{barticle}
\endbibitem

\bibitem[\protect\citeauthoryear{Ma et~al.}{2012}]{MaEtal:12EPES}
\begin{barticle}
\bauthor{\bsnm{Ma}, \binits{J.}},
\bauthor{\bsnm{Wang}, \binits{X.}},
\bauthor{\bsnm{Zhang}, \binits{Y.}},
\bauthor{\bsnm{Yang}, \binits{Q.}},
\bauthor{\bsnm{Phadke}, \binits{A.G.}}:
\batitle{{A Novel Adaptive Current Protection Scheme for Distribution Systems
  with Distributed Generation}}.
\bjtitle{International Journal of Electrical Power \& Energy Systems}
\bvolume{43}(\bissue{1}),
\bfpage{1460}--\blpage{1466}
(\byear{2012})
\doiurl{10.1016/j.ijepes.2012.07.024}
\end{barticle}
\endbibitem

\bibitem[\protect\citeauthoryear{Meliopoulos
  et~al.}{2017}]{Meliopoulos&etal:17TPS}
\begin{barticle}
\bauthor{\bsnm{Meliopoulos}, \binits{A.P.S.}},
\bauthor{\bsnm{Cokkinides}, \binits{G.J.}},
\bauthor{\bsnm{Myrda}, \binits{P.}},
\bauthor{\bsnm{Liu}, \binits{Y.}},
\bauthor{\bsnm{Fan}, \binits{R.}},
\bauthor{\bsnm{Sun}, \binits{L.}},
\bauthor{\bsnm{Huang}, \binits{R.}},
\bauthor{\bsnm{Tan}, \binits{Z.}}:
\batitle{Dynamic state estimation-based protection: Status and promise}.
\bjtitle{IEEE Transactions on Power Delivery}
\bvolume{32}(\bissue{1}),
\bfpage{320}--\blpage{330}
(\byear{2017})
\doiurl{10.1109/TPWRD.2016.2613411}
\end{barticle}
\endbibitem

\bibitem[\protect\citeauthoryear{Liu et~al.}{2017}]{LiuEtal:17TPD}
\begin{barticle}
\bauthor{\bsnm{Liu}, \binits{Z.}},
\bauthor{\bsnm{Su}, \binits{C.}},
\bauthor{\bsnm{Høidalen}, \binits{H.K.}},
\bauthor{\bsnm{Chen}, \binits{Z.}}:
\batitle{{A Multiagent System-Based Protection and Control Scheme for
  Distribution System With Distributed-Generation Integration}}.
\bjtitle{IEEE Transactions on Power Delivery}
\bvolume{32}(\bissue{1}),
\bfpage{536}--\blpage{545}
(\byear{2017})
\doiurl{10.1109/TPWRD.2016.2585579}
\end{barticle}
\endbibitem

\bibitem[\protect\citeauthoryear{Mahat et~al.}{2011}]{MahatEtal:11TSG}
\begin{barticle}
\bauthor{\bsnm{Mahat}, \binits{P.}},
\bauthor{\bsnm{Chen}, \binits{Z.}},
\bauthor{\bsnm{Bak-Jensen}, \binits{B.}},
\bauthor{\bsnm{Bak}, \binits{C.L.}}:
\batitle{{A Simple Adaptive Overcurrent Protection of Distribution Systems With
  Distributed Generation}}.
\bjtitle{IEEE Transactions on Smart Grid}
\bvolume{2}(\bissue{3}),
\bfpage{428}--\blpage{437}
(\byear{2011})
\doiurl{10.1109/TSG.2011.2149550}
\end{barticle}
\endbibitem

\bibitem[\protect\citeauthoryear{}{2019}]{IEEEStd37112-2018}
\begin{botherref}
Ieee standard for inverse-time characteristics equations for overcurrent
  relays.
IEEE Std C37.112-2018 (Revision of IEEE Std C37.112-1996),
1--25
(2019)
\doiurl{10.1109/IEEESTD.2019.8635630}
\end{botherref}
\endbibitem

\bibitem[\protect\citeauthoryear{Jain et~al.}{2019}]{JainEtal19TPD}
\begin{barticle}
\bauthor{\bsnm{Jain}, \binits{R.}},
\bauthor{\bsnm{Lubkeman}, \binits{D.L.}},
\bauthor{\bsnm{Lukic}, \binits{S.M.}}:
\batitle{Dynamic adaptive protection for distribution systems in grid-connected
  and islanded modes}.
\bjtitle{IEEE Transactions on Power Delivery}
\bvolume{34}(\bissue{1}),
\bfpage{281}--\blpage{289}
(\byear{2019})
\doiurl{10.1109/TPWRD.2018.2884705}
\end{barticle}
\endbibitem

\bibitem[\protect\citeauthoryear{Cover and Thomas}{2006}]{Cover&Thomas:book}
\begin{bbook}
\bauthor{\bsnm{Cover}, \binits{T.M.}},
\bauthor{\bsnm{Thomas}, \binits{J.A.}}:
\bbtitle{Elements of Information Theory. (2nd Ed.)}.
\bpublisher{Wiley-Interscience},
\blocation{Hoboken, New Jersey}
(\byear{2006}).
\burl{https://resolver.ebscohost.com/Redirect/PRL?EPPackageLocationID=2106155.274641.21607325&epcustomerid=s9001366}
\end{bbook}
\endbibitem

\bibitem[\protect\citeauthoryear{Gersho and Gray}{1992}]{Gersho&Gray:book}
\begin{bbook}
\bauthor{\bsnm{Gersho}, \binits{A.}},
\bauthor{\bsnm{Gray}, \binits{R.M.}}:
\bbtitle{Vector Quantization and Signal Compression}.
\bpublisher{Kluwer Academic Publishers},
\blocation{Boston}
(\byear{1992}).
\burl{https://resolver.ebscohost.com/Redirect/PRL?EPPackageLocationID=7191.30710607.5628611&epcustomerid=s9001366}
\end{bbook}
\endbibitem

\end{thebibliography}

\end{document}